\documentclass[runningheads]{llncs}

\usepackage[english]{babel}
\usepackage[utf8x]{inputenc}
\usepackage[T1]{fontenc}

\usepackage[a4paper,top=3cm,bottom=2cm,left=3cm,right=3cm,marginparwidth=1.75cm]{geometry}

\usepackage{amsmath,amssymb,amsfonts}

\usepackage{graphicx,tikz}
\usetikzlibrary{tikzmark,arrows.meta,positioning,arrows,shapes,backgrounds,calc,patterns}
\usepackage{multirow}
\usepackage{pgf}
\usepackage[clock]{ifsym} 

\usepackage[colorlinks=true, allcolors=blue]{hyperref} 

\usepackage[ruled,vlined]{algorithm2e}
\usepackage{algorithmic}

\usepackage[ruled]{algorithm2e}

\newcommand\Hilbert[1]{{\cal H}^{#1}}
\newcommand\bra[1]{\langle #1|}
\newcommand\ket[1]{|#1\rangle}
\newcommand\braket[2]{\langle #1|#2\rangle}
\newcommand\reach[1]{\stackrel{#1}{ \longrightarrow }}
\newcommand\rreach[1]{\stackrel{#1}{ \longleftarrow }}

\newcommand\struct[1]{{\bf #1}}
\newcommand\QPN{\struct{Q}}
\newcommand\SysPN{\struct{S}}
\newcommand\Net{\struct{N}}
\newcommand\defqip{{\bf defqp}}

\newcommand\set[2]{\{{#1}\vert{#2}\}}
\newcommand\multiset[1]{\mathbb{N}^{#1}}
\newcommand\preset[1]{{^\bullet}{#1}}
\newcommand\postset[1]{{#1}{^\bullet}}

\newcommand\Minus[1]{{#1}^{-}}
\newcommand\Plus[1]{{#1}^{+}}
\newcommand\Fminus{\Minus{F}}
\newcommand\Fplus{\Plus{F}}
\newcommand\sqrttwo{\sqrt{2}}

\newcommand\sqrtsix{\sqrt{6}}
\newcommand\sqrthalf{\frac{1}{\sqrttwo}}
\newcommand\WF{WF}
\newcommand\PGen{\mathbb{P}}
\newcommand\TGen{\mathbb{T}}
\newcommand\reachset{\mathbb{M}}
\newcommand\reachgraph[1]{G_{#1}}
\newcommand\rategraph[1]{G_{#1}}
\newcommand\ratematrix[1]{R_{#1}}

\newtheorem{convention}{Convention}

\title{How to Bake Quantum into Your Pet Petri Nets and Have Your Net Theory Too}
\author{Heinz W. Schmidt
}

\institute{Computer Science, RMIT University, Melbourne,
  Australia\\
  {\tiny orcid: 0000-0001-6278-4793 \\
    Submitted to and accepted by SummerSOC 21 (summersoc.eu); final version to appear in Springer CCIS
  }
  }

\begin{document}
\maketitle

\begin{abstract}

Petri nets have found widespread use among many application domains,
not least due to their human-friendly graphical syntax for the
composition of interacting distributed and asynchronous processes and
services, based in partial-order dependencies and concurrent
executions. Petri nets also come with abstract semantics, and
mathematical methods for compositional synthesis, structural checks
and behavioural analysis. These have led to the use of various kinds
of nets for real-time, distributed and parallel programming languages,
software and services systems, with a view to their interfaces and
interaction protocols. These affordances make Petri nets invaluable
for distributed software architecture approaches focused on
components, their mutual dependencies and environment-facing
interactions.

Quantum computing -- and in particular quantum software engineering --
is in its infancy and could benefit from the accumulated insights of
software architecture research and of net theory, its methods, and its
applications.

In this paper, we establish a connection between Petri nets and
quantum systems, such that net theory and the component architecture
of nets may help in the synthesis and analysis of abstract software
models and their interface protocols in hybrid classical-and-quantum
programming languages and services systems. We leverage some insights
from net formalisms for software specification for a versatile recipe
to bake quantum into extant Petri net flavours, and prove universality
and compositionality of Petri nets for quantum programming.

\end{abstract}

\noindent{}{\bf Keywords:} 
Component software,
Compositionality,
Petri nets,
Software architecture,
Quantum Petri nets,
Quantum software engineering,
Quantum computing,
Model-checking and simulation,
Stochastic Petri nets


%
%
%
%
%

\section{Introduction}
\label{sec:intro}

Petri nets (PNs) have found widespread use among many application
domains.  Their graphical syntax is tailored for systems exhibiting
concurrency, parallelism and often wide distribution. Their use across
different levels and scales from high-level software systems
requirements through low-level chip architecture to chemical reactions
has assisted their far-flung applications and made PN diagrams a
lingua franca for expressing concurrency directly in its interaction
with causal partial ordering and mutual exclusion
\cite{schmidt2019:PetriNetsNext,reisig1985:PetriNetsIntroduction}.
Petri nets empower systems and services analysts, architects and
developers alike to engage in informal dialogues with each other and
with users, over more or less formal, but accessible, diagrams, about
diverse and complex systems issues including extra-functional
properties such as liveness, reliability and performance.

%
Petri nets also come with an abstract semantics, and mathematical
methods for composition (gluing nets together in different ways) and
for structural and behavioural analysis using the system and services
architecture, i.e., its components, their interrelation and
interaction with their external environment. The {\em unmarked net
  structure} itself is amenable to analysis and transformation,
independent of a specific initialisation in a {\em system net} with a
selected initial marking.  The structure represents the behaviour
rules for all possible initial markings; the system net all runs
connected with the initial marking and possible in that net
structure. The preparation of the initial state remains outside the
model. Net behaviour is observer-independent: a {\em run} of the net
consists of partially ordered occurrences of events tracing, or
simulating, the Petri net token game, in which tokens are
redistributed from one reachable marking to another. Each marking
represents an {\em alternative world} or {\em configuration}. The
events of a run may be {\em sequential}, when one depends causally on
the result of another, or {\em concurrent and spatially distributed},
when they are mutually independent. A set of runs may also reflect
{\em choices}, which may non-deterministically or probabilistically
evolve to different markings. All possible strict sequential event
orders that respect their partial order in a run are {\em conceivable
  observations} of the given run, whether placed on a timeline or not.
Moreover, the system net behaviour can be analysed and transformed
both in terms of an algebra of matrices and the marked net structure
itself.
Thus, an initial marking of the net system provides resource
constraints on top of those set by the net structure. This empowers
modellers to determine many crucial characteristics of a time-less
synchronisation defined solely by the causal dependency and
concurrency of underlying net structure. Different logics have been
used for model-checking in terms of system Petri nets and their
behaviour. Industry-strength model-checkers exist for years capable of
handling hundreds of millions of states (cf. e.g.,
\cite{heiner2013:MARCIEModelChecking,amparore2018:EfficientModelChecking}),
often using special on-the-fly methods for dealing with the 100s of
millions of states without storing transition matrices. The semantics
of PNs has led to their use as formal models for real-time,
distributed and parallel programming languages, for software systems
and services. These affordances make Petri nets invaluable for
distributed software and services architecture approaches focused on
components, their mutual dependencies and environment-facing
interactions.

Quantum computing -- and in particular quantum software engineering --
is in its infancy and could benefit from the accumulated insights of
software architecture research and of net theory, their methods, and
their applications.
Quantum cryptography is already a viable business; so are niche
applications of a number of quantum computing cloud services accessing
the first real quantum computers by IBM and Google with a moderate
number of qubits
\cite{steffen2011:QuantumComputingIBM,castelvecchi2017:IBMQuantumCloud}. Microsoft
and others aim to accelerate the development of quantum software with
breakthroughs expected in the application of quantum computers to
modelling and simulation of real quantum systems that are currently
beyond the reach of supercomputers. These include quantum chemistry
and pharmaceutics, advanced materials, quantum neurophysiology and
others.  Distributed quantum computing is making rapid
advances. Research in optical networks and photonics has made the
quantum internet \cite{kimble2008:QuantumInternet}
feasible. Experimental physics have also demonstrated teleportation of
coherent quantum states experimentally, over many tens of kilometres
and from a ground station on earth to a satellite in orbit
\cite{popkin2017:ChinaQuantumSatellite}. Networked quantum computing
is poised to revolutionise distributed computing and lead to orders of
magnitude increases in network speeds and bandwidth.

However, advances in distributed algorithms and software architecture
for hybrid, i.e., classical-quantum and hardware-software, systems are
far from mainstream software engineering.  Currently the field of
quantum software engineering requires a blend of physics, applied
mathematics and theoretical computer science knowledge. Its models and
methods are inaccessible to most practitioners and academics in
distributed systems analysis, architectural design, software
development and testing.  An exposition of quantum principles in
reasonably widely applied diagrammatic software models and programming
language constructs, like those related to Petri nets, has appeal. For
research, teaching and practice, especially in parallel and
distributed quantum software engineering, PNs could potentially make a
significant difference to practitioners, if quantum were accessible
for different kinds of Petri nets. This requires an orthogonal weaving
of quantum into PNs with a clear separation of concerns of classical
parallelism with its preexisting exponential state space explosion and
the extra quantum-specific parallel features, clearly separating
concurrency and quantum -- while hiding complex vector spaces and
their advanced matrix methods.


%
{\em Overview:} This paper establishes a connection between Petri nets
and abstract quantum systems. We weave quantumness into selected
classes of Petri nets, while separating the concerns for (a) net
structure and (b) non-local and acausal quantum characteristics.
The paper starts with an informal introduction to Quantum Petri Nets
(QPNs) aiming to minimise any vector space knowledge.
We then give a formal definition and show how variations in the
underlying classical nets can be kept separate from quantum
characteristics, yet uniquely extend to QPNs.
The main result shows that Quantum Petri nets (QPNs) can represent any
universal gate set, in fact any circuit in the circuit model of
abstract quantum computation. We also prove a novel compositionality
results.  (Proofs and other technical material are provided as
supplementary material on the web but are here included only for the
convenience of the review process.)
In a section on related and future work we contrast the novel approach
with others, and sketch future research avenues.

\section{A Gentle Introduction to Quantum Petri Nets}
\label{sec:gentle}

Before we introduce QPNs formally and connect them with the
mathematics of quantum information processing, we briefly look at some
simple examples to guide our choices in the design of these
models. Beside addressing PN researchers, this introduction is written
with a community of net practitioners, software developers and applied
informaticians in mind. Many in this community will only be fleetingly
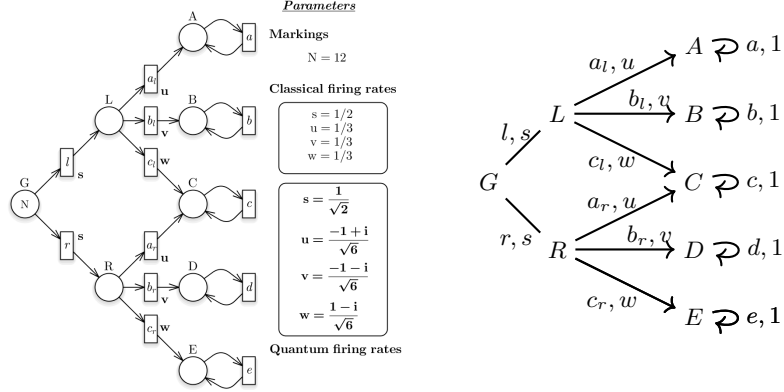
\begin{figure}[htb]
\centering
\includegraphics[width=0.35\textwidth]{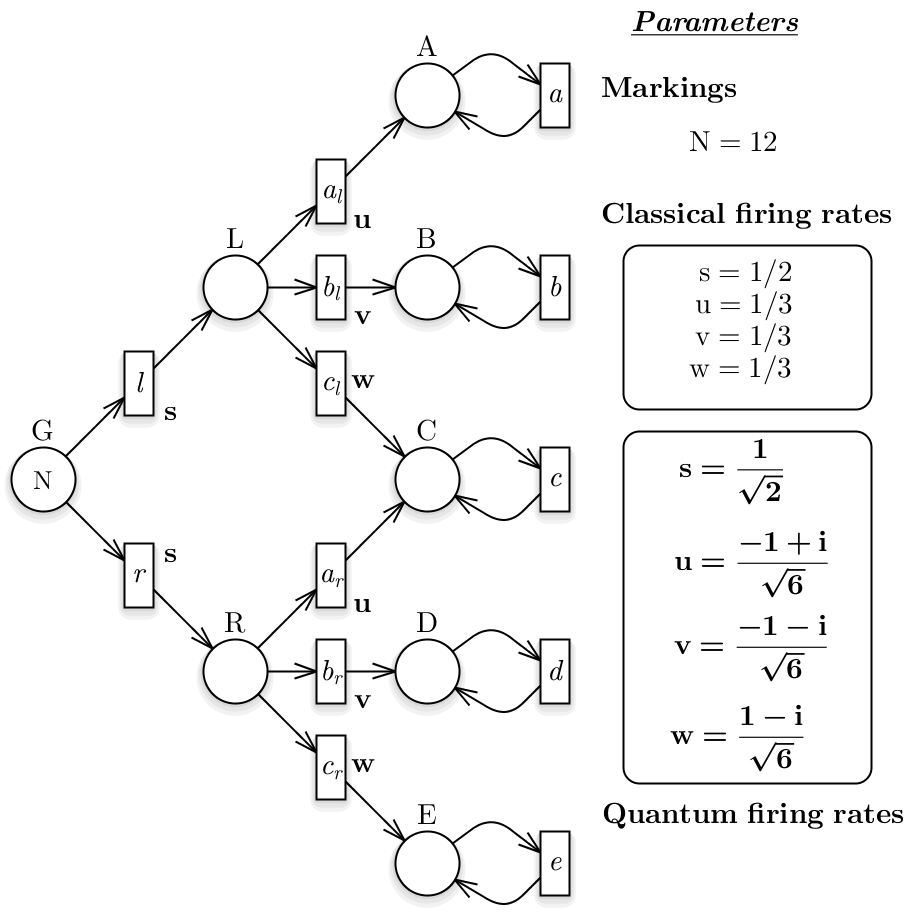}
~~~~~
\raisebox{0.8cm}{
\begin{tikzpicture}[scale=0.9]
                                       \draw(5,8) node(A){$A$};
                       \draw(3.8,7.7) node{$a_l,u$};
                \draw(2.4,6.7) node{$l,s$};
                        \draw(4.35,7.2) node{$b_l,v$};                       
       \draw(3,7) node(L){$L$};        \draw(5,7) node(B){$B$};
                       \draw(3.8,6.25) node{$c_l,w$};
\draw(2,6) node(G){$G$};               \draw(5,6) node(C){$C$};
                       \draw(3.8,5.65) node{$a_r,u$};
                       \draw(4.35,5.2) node{$b_r,v$};                       
       \draw(3,5) node(R){$R$};        \draw(5,5) node(D){$D$};
                \draw(2.4,5.1) node{$r,s$};
                       \draw(3.8,4.2) node{$c_r,w$};                
                                       \draw(5,4) node(E){$E$};
\draw[thick,->] (G) -- (L) -- (A) edge[loop right]node{$a,1$} (A);
\draw[thick,->] (G) -- (R) -- (E) edge[loop right]node{$e,1$} (E);
\draw[thick,->] (L) -- (B) edge[loop right]node{$b,1$} (B);
\draw[thick,->] (L) -- (C) edge[loop right]node{$c,1$} (C);
\draw[thick,->] (R) -- (C);
\draw[thick,->] (R) -- (D) edge[loop right]node{$d,1$} (D);
\draw[thick,->] (R) -- (E) edge[loop right]node{$e,1$} (E);
\end{tikzpicture}
} 

\caption{\label{fig:DoubleSlit} {\em Quantum interference in a simplified particle system model of the double-slit experiment.}
\newline
{\em Left:} For any positive integer $N$, the Place-Transition net
(PTN) carries $N$ tokens on place $G$ only, as its initial marking.
Every transition can fire at least once. That means, the net is live
($L_1$-live). Moreover, one of the transitions $a$ through $e$ will
eventually keep firing (there is no dead marking). When weighting its
transitions with classical rates, the PTN becomes a Stochastic Petri
net (SPN) of the bullet scattering. With quantum rates, it is a
Quantum PTN (QPTN) representing a photon scattering
in a simplified double-slit experiment model. Rates not shown default
to $1$.
\newline
{\em Right:} The weighted reachability relations for N=1 are shown as
a reachability graph. With edge weights, it becomes the weighted
reachability graph of the SPN or QPTN to the left.
All PTN states remain reachable in the SPN.  For the SPN, markings in
$A,B,D$ and $E$ are equiprobable at $1/6$. $C$ attracts tokens from
both slits, hence with twice the probability, $1/3$.
For the QPN however, the probability of reaching $C$ is zero, as the
complex rates cancel each other.  Markings in $A,B,D$ and $E$ are
equiprobable, here $1/4$.  Since $C$ is not reachable in the $QPTN$,
transition $c$ is dead (and the QPTN is not $L_1$-live).
}
\end{figure}

familiar with the mathematics of quantum information processing, if at
all. Therefore, this section is an attempt to explain the basic
principles only in terms of nets and their reachability, wherever
possible. We try to tie principles such as superposition,
entanglement, tunnelling and teleportation directly to the action of
token flow in nets well-known to many in this community.
%
%
In the words of Petri himself, we aim to ``raise the entertainment
value from negative to zero'' \cite{petri1980:IntroductionGeneralNet}
-- at least for this gentle introduction. Where they are needed,
net-theoretic terminology and notation follow that of Reisig
\cite{desel2015:ConceptsPetriNets} widely used in teaching Petri nets
to undergraduate students and practitioners alike. Readers feeling at
home in both the ``mathematical engine room'' of net theory and
quantum information theory, may glance at the figures of this section,
select relevant text explaining the examples and move straight on to
the formalisation of QPNs.
%

In a nutshell, given a classical PN, the corresponding QPN associates
{\em complex-number rates with transition firings, including with
  their inseparable firing as an ensemble of mutually concurrent
  transitions}.

\subsection{Transition Firing with Amplitudes}
\label{sec:amplitudes}

The system net of Fig. \ref{fig:DoubleSlit} uses $N$ tokens ($N$ being
a positive integer) and classical rates to model the outcome of firing
bullets through two slits. Or it uses quantum rates for modelling
photon scattering through the two slits.
With classical rates, the resulting Stochastic Petri Net (SPN)
describes a gun loaded with $N$ bullets (in $G$). With transitions $l$
and $r$, the gun randomly sends a bullet through either the left ($L$)
or right ($R$) slit. For each slit, a scattering ($a,b$ or $c$) occurs
to three of five possible detectors ($A$ through $E$). The location of
a single bullet can then be found always, by firing the respective
transition $a$ through $e$. For these final transitions, the number of
transition firings per time unit can measure the number of bullets
reaching the respective place, if $N>1$.  The probability of a single
bullet of $N$ hitting place $A,B,D$ or $E$, respectively, is
$su=sv=1/6$, i.e., the product of transitions weights on the
respective path from $G$. $C$ is reachable either by the firing
sequence $lc_l$ or $ra_r$, i.e., with probability $sw+su=1/2\times 1/3
+ 1/2 \times 1/3=1/3$.  The token game played on the system
Place-Transition net (PTN) and SPN (to the left) and its reachability
graph (RG) or weighted RG (to the right) reflect the same reachability
relation. As the sum of non-zero probabilities is non-zero for any
SPN, all reachable states have non-zero probability, however small---
assuming that transitions have non-zero probability in the first
place. In our example, like in the underlying PTN, all transitions
$a-e$, including $c$, are enabled at least in some marking. The system
is, what is called in net theory, $L_1$-live.  Moreover, there is no
reachable dead marking, as every token will eventually arrive in one
of the detector places and the corresponding transition will then keep
firing.
Like for its SPN sibling, the quantum firing rates of the Quantum PTN
(QPTN) to the left weight its firings and hence its token game. This
QPTN abstracts from a simple quantum-physical double-slit
experiment. Photons are beamed through two slits and are eventually
detected. In the real experiment, unsurprisingly the light shows an
interference pattern confirming its wave character for very large
numbers of photons ($N$) and any number of runs of the same
setup. However surprisingly to physicists about a century ago, for
repeated measurements of single photons (here modelled by $N=1$) the
same wave pattern appears in the probabilities of photons arriving in
those detector positions, given a sufficiently large number of test
runs. The above system QPTN explains a simple cancellation by
complex-number rates. Amplitude probabilities exclude $C$ from any
marking for all $N\geq 1$.  No token will ever be detected in $C$.
Many details pertaining to a real physical quantum system are
abstracted from our model: for example, in reality, photons bounce
back and force. While complex rates cancel, all tokens make their way
to one of the remaining four detectors with equal probability.

For a quantum firing rate, one uses a complex number $c=a+bi$ and
follows the paths of the net token game, forming products and sums, in
a similar way to that for the SPN above.
In this paper, we do not wish to go into mathematical physics and wave
interpretations of complex rates. Suffice to say, that complex numbers
allow compact characterisations of real classical mechanical waves and
also of complex waves appearing in quantum mechanics.
%
A complex number is also called {\em amplitude}. Its modulus
$|c|=|\sqrt{a^2+b^2}|$ measures its real magnitude, its modulus
squared $|c|^2=a^2+b^2$ a real probability, possibly after some
normalisation in the context of other complex numbers.  So in the
example, we have, $su=\frac{-1+i}{\sqrttwo\sqrtsix}$,
$sv=\frac{-1-i}{\sqrttwo\sqrtsix}$, and
$sw=\frac{1-i}{\sqrttwo\sqrtsix}$.  Since amplitudes can cancel,
$|sw+su|^2=0$. No token can reach $C$ in the system QPTN, whether
calculated on the net structure or the weighted reachability steps in
its token game and visualised in the RG to the right. Moreover, that
is the case, however many experiments are run, and however large
$N\in\mathbb{N}$ is chosen to begin with. Hence transition $c$ is dead
and the QPTN is not live for any $N$.  On the other
hand, a short calculation shows that $A, B, D$ and $E$ are
equiprobable.  We normalise each their four amplitudes with their sum
total magnitude $|m|=|su|+2|sv|+|sw|=|\sqrt{4}/\sqrt{6}|$. Place $A$,
for instance, is therefore reached with probability
$(|su|/|m|)^2=|su|^2/|m|^2=\frac{1}{6}/\frac{4}{6}=1/4$.
%
%
%

\subsection{Superposition States and Measurements Over Nets}
\label{sec:superposition}

Let $\reachset_0$ denote the {\em reachability set}, i.e., the set of
reachable markings, of the underlying PN with initial marking
$M_0$. For example, in Fig. \ref{fig:DoubleSlit}, $\reachset_0$ is the
set of nodes in the RG on the right.  A {\em superposition state} $a$
({\em superposition} for short) is a function
$\reachset_0\reach{a}\mathbb{C}$ that assigns a complex number to each
marking in $\reachset_0$.  The set of all superpositions of the
reachability set $\reachset_0$ is also denoted as $\mathbb{A}_0$ and
called the {\em span of the QPN}. We call $a(m)$ the {\em probability
  amplitude} of marking $m\in\reachset_0$ in $a$ and write $a_m$
instead of $a(m)$ for brevity. We also denote $a$ algebraically as a
weighted sum of its markings, each placed in a special bracket:
$a_1\ket{m_1}+\cdots+a_n\ket{m_n}$, omitting terms with weight $0$.
For convenience in examples, we avoid the brackets, when there is no
risk of confusion.
%
For real-valued amplitudes, the superposition represents classical
{\em alternative worlds}. The trivial superposition $\ket{m}$ (i.e.,
$a_m=1$ and other $a_{m'}=0$) represents a single marking. We
liberally use familiar algebraic manipulating of superpositions (and
defer the formal treatment to Sec. \ref{sec:formal}).
%
%
%
%

The state of a quantum system is not directly observable. Observation
requires a measurement of some sort.  A superposition expresses
Heisenberg's uncertainty principle inherent in quantum systems and
making measurement outcomes uncertain.  Once measured, the subsystem
will be in a certain classical state. Physicists speak of the {\em
  collapse} of the wave function.
%
Measurement in a real test setup requires a physical instrument, which
is ultimately a quantum system itself, interfering with the system
under test (commonly abbreviated SUT in software engineering). For
example, a software-defined nanorobotic system may consist of a
software-controlled laser measuring and controlling some nanorobotic
material in superposition. Light beams sent from the laser change the
energy levels and conductivity of particles making up that
material. They affect its properties and shape.
Any test effect before the ultimate collapse can be regarded as a
transition in the SUT from the superposition $a$ prior to the test to
another superposition $b$ immediately before the ultimate collapse.
That transition is typically forgetful and hence irreversible, because
it abstracts from many characteristics of the system not of interest.
The ultimate collapse remains unpredictable.

For QPNs, the result of the ultimate collapse is a marking $m$.
That measurement and collapse may be restricted to a part of the
system only, not entangled (see below) with the rest of the
system. Then collapse may be restricted to the subsystem state.
%
So, having modelled the test setup in terms of its effect on the SUT
by a transition from $a$ to $b$, the modeller can now inspect the
amplitude of a marking $m$ in $b$ (in the model) to determine the
probability of that outcome (in the actual system) as $|b_m|^2$, if
$b$ is {\em normal}%
\footnote{Physical model-realism may require operations on quantum
systems at any scale and hence resulting in superpositions at any
scale. This may require re-normalisation.}.
A superposition $a$ is defined as normal if
$a=a/|a|$, where $|a|$ is the real-valued norm of the superpositon
defined by $|a|=\sqrt{\sum_m |a_m|^2}$.

\begin{figure}[htb]
\centering
\includegraphics[width=0.3\textwidth]{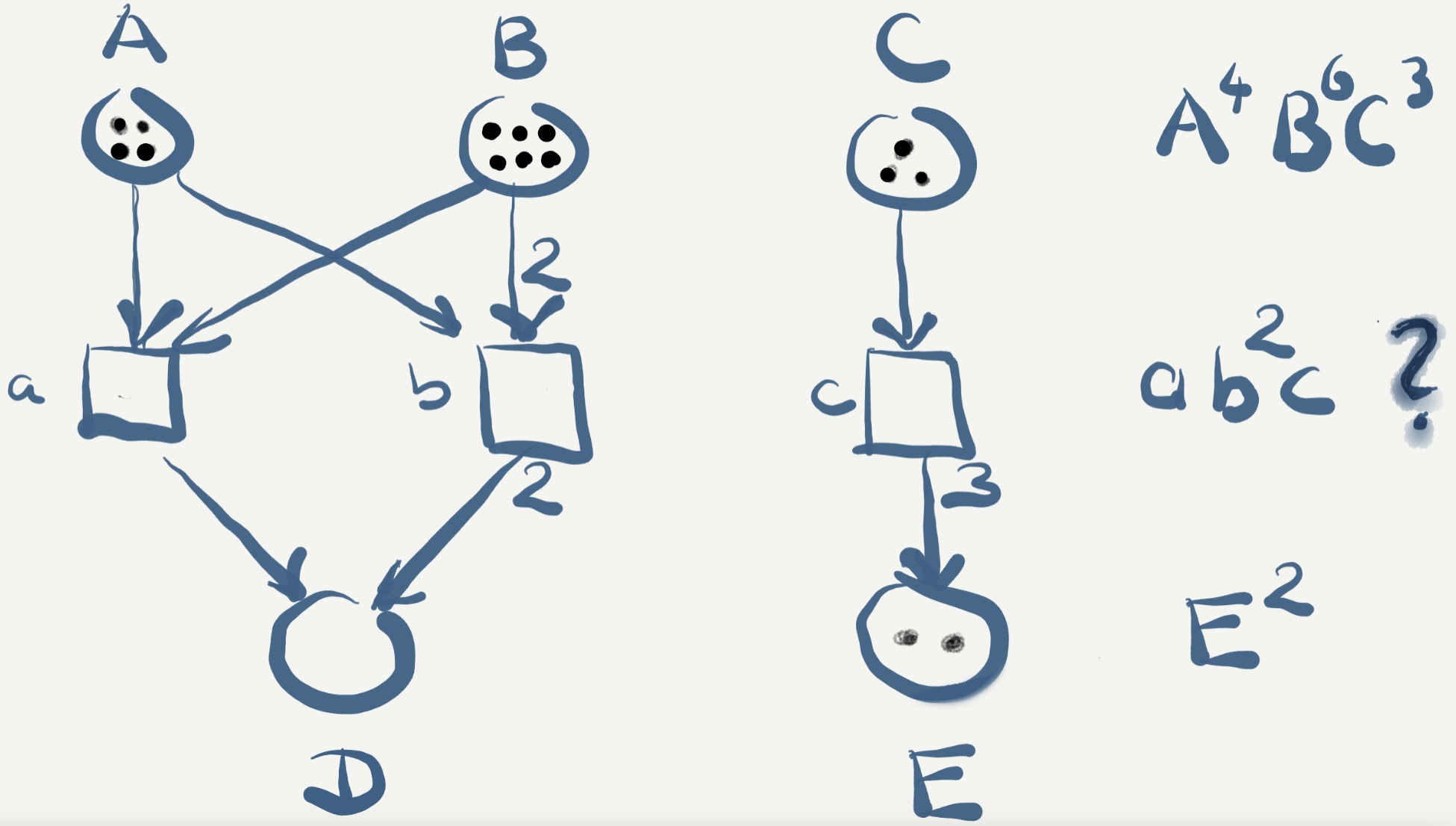}
\includegraphics[width=0.32\textwidth]{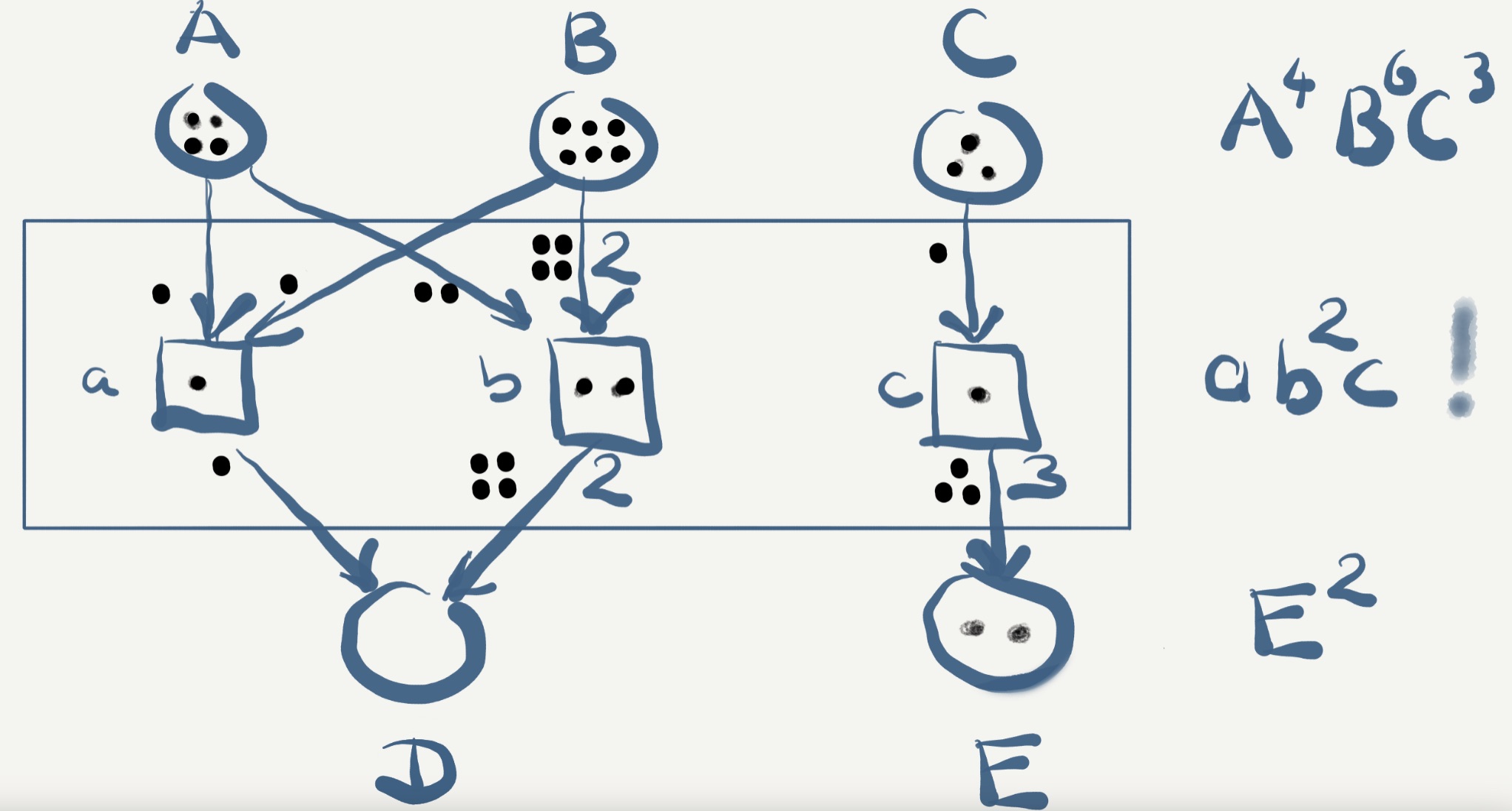}
\includegraphics[width=0.3\textwidth]{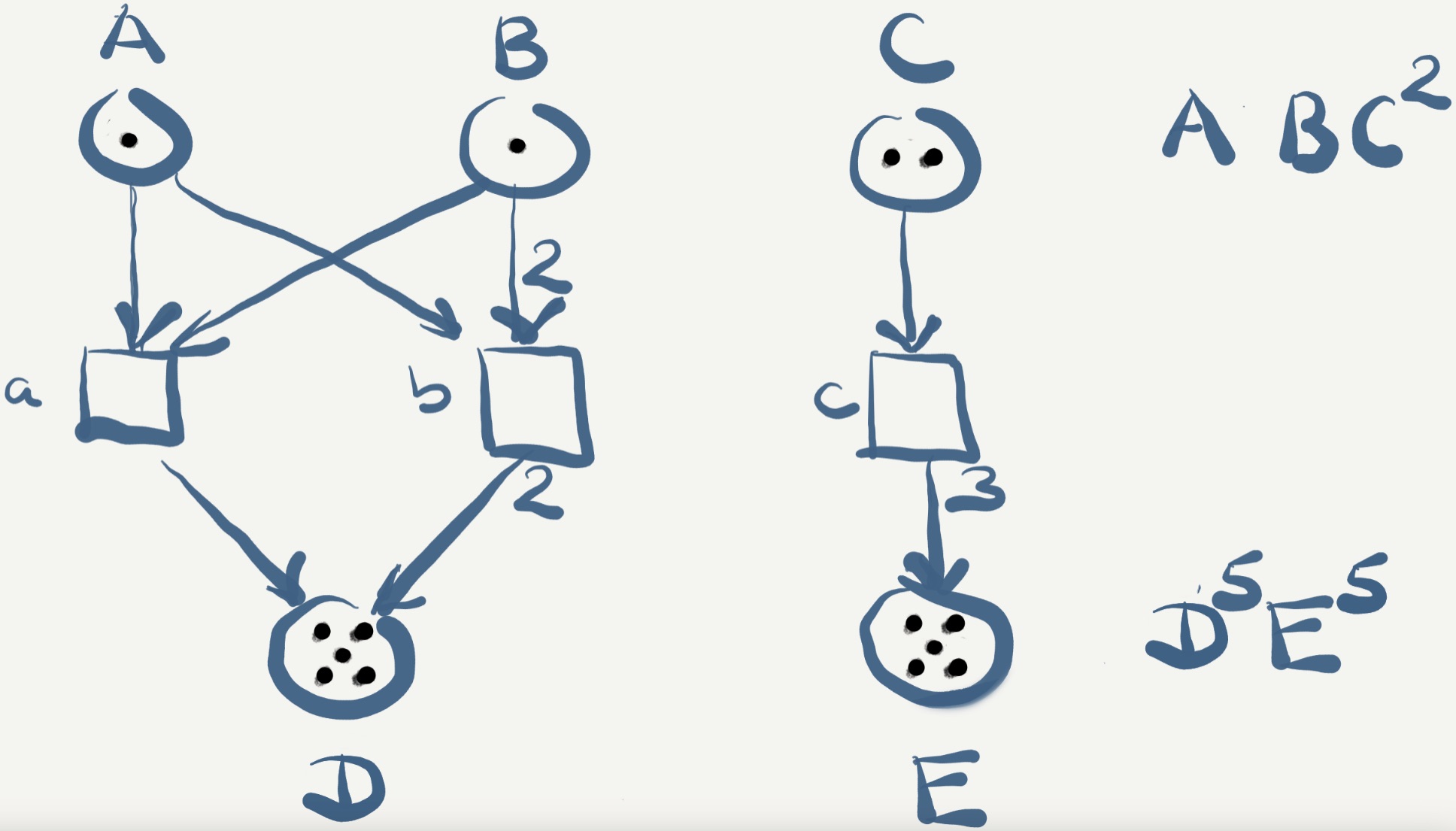}
\begin{center}
\caption{\label{fig:concurrence} {\em Concurrence enabling and firing.}
\newline
{\em Left:} The marking $A^4B^6C^3E^2$ enables all transitions $a,b$
and $c$ in the concurrence $ab^2c$. Does it enable them in the
required number?
{\em Middle:} The net flow multiplicities (arc
weights), which define the input and output markings of a single
transition, are lifted to input and output markings of a
concurrence. These determine the interference-free coupled firing of the
multiset. Here the input marking is $A^3B^5C$. Likewise, the output
marking depicted respects multiplicity, here $D^5E^3$.
{\em Right:} Thus the marking $ABC^2D^5E^5$ can be reached by firing
$ab^2c$: $A^4B^6C^3E^2\reach{ab^2c}ABC^2D^5E^5$. As usual, this joint
firing subtracts tokens and adds tokens in a single atomic step. The
overall classical effect on the marking is the same as firing
subconcurrences in any sequence, no new marking is reachable. However,
with entanglement, the transition multiset may fire any way: together,
individually or anything in between.}
\end{center}
\end{figure}

%


\subsection{Entangled Transitions and States}
\label{sec:entanglement}

Quantum-mechanically, entanglement requires superposition of {\em
  multiple interacting subsystems}.  For example, two or more
independent molecules may each be in a superposition of conformations,
when ``swinging'' on a double-bond hinge or ``rotating'' on a
single-bond pivot. These conformations may be frequency-coupled
between the molecules, due to the surrounding electronic structure,
i.e. the distribution of a number of electrons they share and the
uncertainty of their position.
In QPNs, entanglement couples\footnote{We use the term space-time
coupling to avoid the term `synchronisation' which is loaded with
temporal meaning, etymologically the old-Greek `khronos' means time
and the prefix `sun' together.} the firing of otherwise concurrent
transitions in space-time. We call the firing of mutually concurrent
transitions a {\em concurrence}\footnote{Concurrences are to
transitions what markings are to places.}. Formally, a concurrence is
a multiset $t$ of enabled and mutually independent transitions, i.e.,
$t\in\multiset{T}$ (the set of multisets over $T$). We think of them
as firing together with some given amplitude.  Like the transition
structure of nets is the dual of its place structure, concurrences can
be regarded as the dual of markings. However, like single transitions,
concurrences define indistinguishable atomic marking updates and must
not be confused with historic attempts in net theory to separate the
beginning and the ending of transitions.  As an abstract example of a
concurrence, consider the concurrence $ab^2c$ in
Fig. \ref{fig:concurrence}. The shown marking $A^4B^6C^3E^2$ has
sufficient tokens for an interference-free firing of any pair of
subconcurrences partitioning and covering the given concurrence. Hence
it can also fire in an atomic {\em single step}. This example also
shows that concurrences do not have to be maximal.

\begin{figure}[htb]
\centering
\raisebox{-1.35cm}{\includegraphics[width=0.35\textwidth]{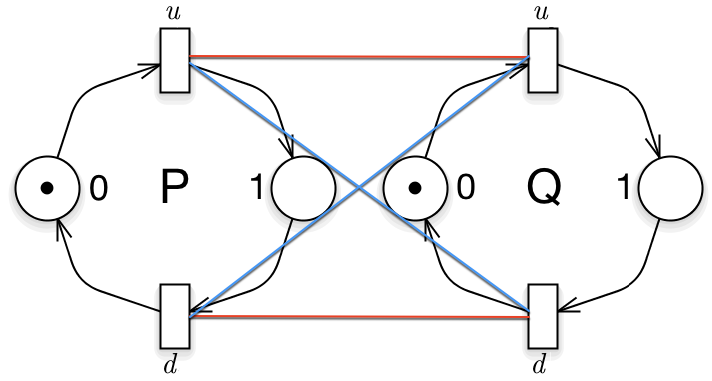}
         }
\raisebox{-1.45cm}{
\begin{tikzpicture}[scale=0.45]


        \draw(2.5,5.7) node[anchor=south](UPQ){$\cdot d$};
\draw(1,4.5) node(DD){00}; \draw(2.5,4.5) node[anchor=south]{$\cdot u$}; \draw(4,4.5) node(DU){$01$};
\draw(-0.35,3) node(UPP)[anchor=east]{$u\cdot$}; \draw(1,3) node[anchor=east]{$d\cdot$}; 
              \draw(4,3) node[anchor=west]{$u\cdot$}; \draw(5.45,3) node(DOP)[anchor=west]{$d\cdot$};
  \draw(1,1.5) node(UD){$10$}; \draw(2.5,1.5) node[anchor=north]{$\cdot d$}; 
                                \draw(4,1.5) node(UU){$11$};
        \draw(2.5,0.3) node[anchor=north](DOQ){$\cdot u$};
        
\draw[blue](1.7,3.7) node {$du$}; \draw[red] (3.3,3.6) node {$uu$};
\draw[red] (1.7,2.4) node {$dd$}; \draw[blue](3.25,2.4) node {$ud$};

\draw[thick,->] (DD) -- (DU);
\draw[thick,->] (DU) -- (UU);
\draw[thick,->] (UU) -- (UD);
\draw[thick,->] (UD) -- (DD);

\draw[thick,->] (DU) edge[bend right=70] (DD);
\draw[thick,->] (DD) edge[bend right=70] (UD);
\draw[thick,->] (UD) edge[bend right=70] (UU);
\draw[thick,->] (UU) edge[bend right=70] (DU);

\draw[red,thick,->]  (DD) edge[bend left=40] (UU);
\draw[red,thick,->]  (UU) edge[bend left=40] (DD);
\draw[blue,thick,->] (UD) edge[bend left=40] (DU);
\draw[blue,thick,->] (DU) edge[bend left=40] (UD);

\end{tikzpicture}
}
$
\begin{bmatrix}
  r_{00}(1)        & r_{01}(\cdot d)  & r_{10}(d\cdot)   & r_{11}(dd) \\
  r_{00}(\cdot u)  & r_{01}(1)        & r_{10}(du)       & r_{11}(d\cdot ) \\
  r_{00}(u\cdot)   & r_{01}(ud)       & r_{10}(1)        & r_{11}(\cdot d) \\
  r_{00}(uu)       & r_{01}(u\cdot)   & r_{10}(\cdot u)  & r_{11}(1)
\end{bmatrix}
$

\caption{\label{fig:entangled} {\em Complex firing rates turn a pair of bits into a pair of qubits}\newline
  {\em Left:} A QPN system with underlying PTN of two unconnected
  bits, $P$ and $Q$, in black. Initially both bits are $0$. The
  possible PTN transitions, in each, flip the respective bit up ($u$)
  or down ($d$). Pairs of mutually concurrent transitions are
  highlighted by coloured lines. QPN rates default to $1$.
  \newline
  {\em Middle:} The reachability graph of the PTN to the left is shown
  in black.  A marking $P_iQ_j$ is abbreviated by juxtaposition $ij$,
  for short (e.g., $01$ for $P_0Q_1$). Likewise, the reachability
  edges are labeled with juxtaposed transition symbols, for short,
  filling in $\cdot$ for no action (NOP) on the respective subnet
  (e.g., $\cdot u$ for $Q_u$). The coloured reachability edges add
  concurrences firing entangled transition multisets (e.g., $uu$).\newline
  {\em Right:} Adjacency matrix $A$ of the QPN RG with rows and
  columns indexed $00$ through $11$. For every reachability step
  (edge) $m\reach{t}m^{\prime}$, we have
  $A_{m^{\prime},m}=r_{m}(t)$. NOP is represented by the unit multiset $1$. }
\end{figure}

Thus, for QPNs, concurrences can be entangled by rating the
inseparability of any of their proper substeps. In fact, it is
possible to acausally force their atomic firing by rating the
individual transition firings with amplitude $0$, while using
amplitude $1$ for the joint firing.  This is a kind of soft mutual
excitement. Inversely, the joint firing could be rated much lower than
the individual firing, as a kind of soft mutual inhibition.  A {\em
  marking-dependent rate function} $r$ is a function $\reachset_0
\times \multiset{T} \reach{r} \mathbb{C}$ assigning complex numbers to
concurrences. We refer to $r(m,t)$ by $r_m(t)$, for brevity, and use
$r_m(t)=1$ as the default.
For example, in Fig. \ref{fig:entangled}, there are four possible
concurrences between the two classical bit PTNs $P$ and $Q$. Here each
concurrence consists of two transitions highlighted by a coloured
line. As rates are not shown explicitly, they default to $1$, and turn
the classical bit pair into the QPN of a qubit pair.
%
%
%
A classical net composition by juxtaposition of their graphs means
figuratively that independent players can move tokens according to the
usual token game, each on their net component. Formally, the usual
reachability of the combined nets is the Kronecker product of the
reachability graphs of the component nets. In the formalisation of
QPNs below (Sec. \ref{sec:formal}) we write this as $P\otimes Q$ (see
e.g.,
\cite{campos1999:StructuredSolutionAsynchronously,donatelli2001:KroneckerAlgebraStochastic}
and advanced books on graph theory and matrices). Classical net
architecture design usually discourages disconnected nets and adds
further constraints to two juxtaposed components.  For example, the
additional constraint $P_u=Q_u$ and $P_d=Q_d$ (gluing the respective
pair of transitions in the juxtaposed nets highlighted by red
lines). This forces the respective players into mutually dependent
moves -- players holding hands, so to speak, in their now locally and
causally synchronised token game. In this example, the two nets
juxtaposed with this particular constraint are behaviourally
equivalent to a single classical bit: the RG of the combined net is
isomorphic to the RG of each single player.
%
%
%
Juxtaposition with transition identification loses the individual
transition firings.  In contrast, a {\em QPN can fire all transitions
  of a concurrence together, individually, or anything in
  between} controlled by the rate function.

Formally, an {\em entangled superposition} (state) is defined as a
superposition that cannot be expressed as a product of the two
component superpositions.  In Fig. \ref{fig:entangled}, for instance,
$\sqrthalf(\ket{01}+\ket{10})$ is an entangled state. In contrast, the
superposition
$\frac{1}{2}
(i \ket{00}
-i \ket{01}
-  \ket{10}
+  \ket{11}) =
\frac{1}{2}
(i \ket{P_0Q_0}
-i \ket{P_0Q_1}
-  \ket{P_1Q_0}
+  \ket{P_1Q_1}) =
\frac{1}{2}
( \ket{P_0} +i \ket{P_1}) \times 
            (i \ket{Q_0} -i \ket{Q_1}) $
is a product of superpositions of the $P$ and $Q$ qubit taken by
themselves. Therefore this superposition is not entangled.
%
%
The so-called Bell states\footnote{ named after
physicist John S. Bell. These states are maximally entangled.}  of a
qubit pair are the four entangled superpositions of the reachability
set of Fig. \ref{fig:entangled}(Middle):
$\sqrthalf(\ket{00}\pm\ket{11})$ and
$\sqrthalf(\ket{01}\pm\ket{10})$. To stay in
$\sqrthalf(\ket{00}\pm\ket{11})$, for instance, $uu$ and $dd$ have to
be rated with probability $1$ (or a complex equivalent) with $u$ and
$d$ rated $0$. Informally, rating the individual transitions in the
coupling both with amplitude $0$ means that the intermediate markings
$01$ and $10$ attract an amplitude of $0$ as result of firing the
concurrence in a single step.

\subsection{Superposition Evolution and Complex Token Game}
\label{sec:evolution}

{\em Complex rates for single transitions and for concurrences allow a
  QPN to evolve the state from superposition to superposition with
  entanglement}. An evolution process starts from a {\em initial
  state}: any superposition is a valid initial state. The QPN process
executes in parallel for all superposed markings and all possible
concurrences enabled in any of them. QPN superposition evolution is
thus a form of OR-parallelism, well-known from parallel and stochastic
process modeling and simulation tools, incl. SPNs.  Each step reaches
another superposition; a measurement is taken in a distinguished
state, for example a final state, or a home state that is both initial
and final. The superposition immediately before the collapse reveals
the probability of measurement outcomes with a precisely defined
uncertainty\footnote{The precision of the mathematics underpinning
quantum mechanical predictions is unrivaled and differs from classical
stochastic models.}.

Recall like in stochastic variants of Petri nets, QPN rates are {\em
  marking-dependent}.  In the RG of an SPN, this permits a modeller to
assign different rates $r_m(t)\not=r_{m^\prime}(t)$ to different
enabling markings for the same transition ($m\reach{t}$,
vs. $m'\reach{t}$).
For QPNs, the step $t$ may be a concurrence. Rating a concurrence with
a single transition ($|t|=1$) is trivial and classical.  Rating the
unit multiset ($|t|=0$), i.e., the NOP process, caters for a {\em
  resting} rate. For example something may change somewhere remote (in
another juxtaposed component) and result in redistribution of
amplitudes, in particular for non-local entanglements.
Any concurrence and its entanglement can now be fully grasped in
terms of transition firings in the net.
For example, consider the RG (black and coloured) in
Fig. \ref{fig:entangled}. Assume we want to compute the amplitude of
marking $\ket{11}$ after a direct reachability step starting from a
given superposition $a$. Given the uncertainty expressed in $a$, there
are many steps of length $1$, that reach $\ket{11}$, i.e. the target
node $11$ in the RG on the right of the figure. To compute the
required amplitude (here $a_{11}$), one forms the weighted sum over
these edges with their source node (marking) amplitudes as the
respective weight, plus the weighted resting rate.
\begin{equation}\label{equ:evolution}
  \begin{array}{lllll}
    a_{00} & :=  a_{00}\times r_{00}(1)       &+~ a_{01}\times r_{01}(\cdot d) &+~ a_{10}\times r_{10}(d \cdot ) &+~ a_{11}\times r_{11}(dd) \\
    a_{01} & :=  a_{00}\times r_{00}(\cdot u) &+~ a_{01}\times r_{01}(1)       &+~ a_{10}\times r_{10}(du)       &+~ a_{11}\times r_{11}(d\cdot) \\
    a_{10} & :=  a_{00}\times r_{00}(u\cdot)  &+~ a_{01}\times r_{01}(ud)      &+~ a_{10}\times r_{10}(1)        &+~ a_{11}\times r_{11}(\cdot d) \\
    a_{11} & :=  a_{00}\times r_{00}(uu)      &+~ a_{01}\times r_{01}(u\cdot)  &+~ a_{10}\times r_{10}(\cdot u)  &+~ a_{11}\times r_{11}(1)     
  \end{array}
\end{equation}
Since node $11$ is reachable only by making a move (a step in one of
the first three terms listed) or by resting (fourth term), the
normalisation of the rates in Equ. \ref{equ:evolution} (line $4$)
results in the sum of their moduli squared equaling $1$, provided that
$a$ is normalised to begin with. Philosophically, the required
probability sum of $1$ (after normalising and converting amplitudes to
probabilities) expresses this principle\footnote{in a blend of
Einstein and Aristotle}: {\em whether something moves or nothing
  happens, the system is in a defined state}. In other words, a
measurement will collapse the state to a defined outcome.

For each marking amplitude $a_m$, a corresponding equation in
Equ. \ref{equ:evolution} captures the direct reachability relation
reaching marking $m$.  Note that according to Equ. \ref{equ:evolution}
the amplitudes of the next superposition are exactly the result of
multiplying row $11$ of the adjacency matrix $A$
(cf. Fig. \ref{fig:entangled}) with $a$ as a vector of marking
amplitudes in the order $00$ through $11$.
Step sequences of different lengths (lenghts of paths in RG) and width
(cardinality of concurrences) represent partially ordered runs. The
net structure of a QPN and various behaviourally equivalent
transformations on the classical underlying net
\cite{berthomieu2018:PetriNetReductions} may serve transformations of
the QPN with different complexity and performance characteristics, yet
resulting in the same final measurement up to an error $\Delta$.

Note, that the rate function does not change the reachability of the
underlying classical net. Therefore, QPNs do not add to the classical
state space explosion already attributable to concurrency and
stochasticity. For example, SPNs already calculate with stochastic
uncertainty in their parallel simulations. However, QPNs add
concurrence edges to their reachability graph filling the equivalent
rate matrix with finely differentiated rates.
A secondary and lesser complication arises from the use of complex
numbers as rates. Each complex number requires two reals and hence
doubles the space and possibly access time compared to SPN
implementations. More importantly, adding classical probabilities
increases them and multiplying them decreases them. Therefore, rates
for classical stochastic nets may converge over increasing numbers of
steps, allowing a modelling tool to prune the number of OR-parallel
branching.  In contrast, complex rates may cancel each other, yet
oscillate forever, with no or little chance of reducing the number of
processes to follow.


\subsection{Rate Function Composition and Acausal Computation}
%
In general, any quantum computational function can be expressed using
QPNs. The causal characteristics of the computation is captured in the
net structure. The acausal characteristics are
\begin{figure}[htb]
\centering
\raisebox{-0.6cm}{\includegraphics[width=0.12\textwidth]{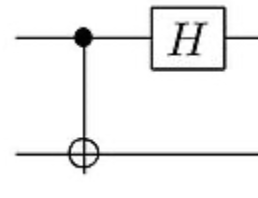}
  }~~
\begin{tabular}{|l|l}
  ~~\defqip~ {\tt (Qubit P,Q).CNot {@} 0} {\bf with}  ~~ & ~~ \defqip~ {\tt (Qubit P).Had {@}} $\sqrthalf$ {\bf with} \\
  ~~~~~{\tt P.0 {\bf rest} {@} 1 }   ~~ & ~~~~~ {\tt P.1 {\bf rest} {@}} $- \sqrthalf$ .\\
  ~ {\tt | P.1 ( Q.u | Q.d ) {@} 1 .}     ~~ & ~
\end{tabular}
\begin{center}
  \caption{
    \label{fig:cnoth}
          {\em Quantum circuits and rate functions.}
    {\em Left:} A quantum circuit diagram with two qubits (horizontal
    lines, say $P$ at top, $Q$ at bottom) and two gates (operating on
    lines left to right). The first gate, a so-called CNOT gate, does
    nothing when $P=0$. For $P=1$ it flips the superposition of
    $Q$. The second gate, a so-called Hadamard gate maps
    $\ket{0}\mapsto \sqrthalf(\ket{0}+\ket{1})$ and $\ket{1}\mapsto
    \sqrthalf(\ket{0}-\ket{1})$.
    {\em Middle:} The amplitude definition of CNOT as rate function
    {\em CNot} over the QPN in Fig. \ref{fig:entangled}.
    {\em Right:} Hadamard gate as rate function {\tt Had} over single qubit $P$ in Fig. \ref{fig:entangled}.
}
\end{center}
\end{figure}

induced by the firing steps and thus direct reachability, in terms of
the overall rate distribution over the net. Below, we show by example
(Fig. \ref{fig:entangled} and \ref{fig:cnoth}), how quantum circuit
diagrams translate to QPNs. For technical details see
Sec. \ref{sec:formal}.

In circuit diagrams, the controlled-not gate, aka CNOT, assigns
amplitudes so as to flip qubit $Q$ if and only if qubit $P$ is $1$,
with the rate function CNot induced by the steps of the underlying net
of the qubit pair $P$ and $Q$. The corresponding QPN amplitude
definition (\defqip) declares {\tt CNot} as a rate function on the
qubits $P$ and $Q$, induced by the steps of the underlying net.  The
default amplitude is $0$ (`{@} 0' follows the rate function
prototype). The {\tt CNot} definition body follows the {\bf with}
keyword and defines marking-dependent rules overriding the default for
firing steps.  With $P.0$ marked the resting amplitude is $1$ (`{\bf
  rest} {@} 1'), i.e., regardless of the marking of $Q$. This means
$CNot_{00}(1)=CNot_{01}(1)=1$.  With $P.1$ marked, the system is
forced to fire one step to a different marking as the resting rate is
$0$ by default for the remaining markings. In a superposition, the
enabling markings have different amplitudes, but for each marking
there is only one enabled transition: $10\reach{Q.u}$ and
$11\reach{Q.d}$. Thus with $P.1$ the rate is
$CNot_{10}(.u)=CNot_{11}(.d)=1$, thus performing the respective flip
with probability $1$. Using parentheses, the two transition amplitude
rules are contracted here into one rule for $P.1$.

Similarly, the so-called Hadamard gate, which operates on the single
qubit, here $P$ in Fig. \ref{fig:entangled}, is given by the rate
function {\tt Had} at amplitude $1/\sqrt{2}$ for all steps bar
$H_{1}(1)=-1/\sqrt{2}$.

%
Each rate function on one or more qubits corresponds to a primitive or
composite gate of quantum computing circuits on those qubits.  QPNs
are not limited to nets representing bits, however. Hence, more
generally, each rate function corresponds to a combined
causal-and-acausal function of the QPN, in which amplitudes regulate
casaulity, concurrency and choice.

\subsection{Teleportation Tunnelling and Hybrid Security Protocols}
\label{sec:teleportation}

Due to the particle-wave character of {\em entangled quantum systems},
their behaviour includes instantaneous synchronisation over large
spatial distances, local-point correlations over temporal distance
reaching into the past and future, and blends thereof in bounded
space-time regions. Quantum-state teleportation experiments have
confirmed qubit entanglement over large distances. For example, qubit
$P$ could be located on Earth while qubit $Q$ is on a satellite in
orbit.  If the entangled state $\sqrthalf(\ket{00}+\ket{11})$ were
distributed with $P$ in space and $Q$ on Earth and a measurement in
space resulted in the outcome of $P_0$ (and thus a collapse of that
state to $P_0Q_0=00$), then instantaneously it can be known that the
qubit $Q$ on Earth is in state $Q_0$. An analogous argument holds for
the alternative outcome $P_1$. Furthermore, symmetrically, the
measurement could be taken on Earth with the same non-local
result. Yet the outcome remains uncertain until a measurement is
taken. Superposition does not only work like a perfect quantum
oracle. It also implies instantaneous consumption
of distributed measurement outcomes.  Moreover, any eavesdropping on
entangled qubits results in a collapse of the superposition, which can
be immediately noticed by the distributed parties. Therefore, quantum
protocols are also poised to offer unparalleled security and safety.

It should be noted, however, that the acausal and non-local
entanglement is not signalling any information between remote
locations.  However, it can be used in combination with classical
message signalling protocols over such distances to leverage the
quantum advantage in protocol security and processing speed.
Using the examples above we have the tools to understand the following
teleportation protocol, which abstractly captures the common structure of a
number of real teleportation protocols that have been conducted over
different distances (tunnelling under the Donube in Germany, offshore
to onshore US, Tibet to space satellite and others, cf.,
e.g. \cite{yanofsky2008:QuantumComputingComputer}).

In this variant of the protocol, Alice and Bob each share one half of
an entangled qubit pair ($A$ and $B$ say), in the Bell state
$\sqrthalf(\ket{00}+\ket{11})$, which results from evolving $\ket{00}$ 
with $(A,B).Bell$ in our QPTN.
They both have agreed on the entanglement but neither has knowledge of
the specific Bell state or its amplitudes.  Alice aims to teleport a
third qubit $C$ to Bob.
%
%
The state of $C$ -- also unknown to Alice and Bob -- can be
represented as the superposition of markings $c\ket{0}+d\ket{1}$. Now
juxtapose the QPTNs in the order $C, A$ and
$B$ with our convention for marking abbreviations $000$ through $111$.
Alice is located in the space station, Bob on Earth.  The entanglement
of $A$ and $B$ is prepared in a third place before the experiment
using the {\tt Bell} function above. Next the separate qubits are sent
to Alice and Bob taking care not to break the entanglement.  This
completes the preparation of the initial three-qubit superposition
$\sqrthalf(c\ket{000}+c\ket{011}+d\ket{100}+d\ket{111})$. Now:
\begin{enumerate}
\item Alice uses {\tt (C,A).CNot; C.Had} (Fig. \ref{fig:cnoth})
  transferring the entanglement from $A$ and $B$ to $A$ and $C$. This
  results in
  $\frac{1}{2}(c(\ket{000}+\ket{011}+\ket{100}+\ket{111})+d(\ket{010}+\ket{001}-\ket{110}-\ket{101}))$
  which can be refactored
  $\frac{1}{2}(\ket{00}(c\ket{0}+d\ket{1})+\ket{01}(c\ket{1}+d\ket{0})+\ket{10}(c\ket{0}-d\ket{1})+\ket{11}(c\ket{1}-d\ket{0}))$
  to show Bob's qubit as a function of any possible marking of Alice's qubit pair.
  \item Next Alice measures her qubits $(C,A)$ and finds them in one
    of the four underlying markings, $00$, $01$, $10$ or $11$ shown above,
    and leaving the amplitudes of $C$ hidden in Bob's formerly entangled
    qubit. Alice sends the two classical bits of that measured marking to
    Bob.
  \item To reveal $C$, Bob applies one of four functions, depending on
    the message sent and arrives in superposition $c\ket{0}+d\ket{1}$
    for $C$. Note that Alice has lost the initial state of $C$ in the
    process.
\end{enumerate}

We view tunnelling as a massive entangling with collective
teleportation. For example, a long sequence of juxtaposed QPNs of
identical system nets can represent a tunnel architecture, with only
moderate causal connections between adjacent QPNs. Now the instances
of the same transition across different positions can be entangled and
teleportation achieved from one end of the tunnel to the other.  Causal
interconnections in the architecture may reinforce entanglement over the
length of the tunnel. Concurrences over instances $t_i$ of the same
transition are powers $t^n$ (ignoring the position).  Firing rates for
these powers may be expressed in terms of the exponent $n$ in one net.
  
\section{Quantum Petri Nets with Variation Points}
\label{sec:formal}

QPNs have standard Petri nets as underlying classical nets. To avoid
ambiguity of similar terminology and notation in nets and quantum
computing, we reconcile basic notation first for nets and their
matrices.  This will also provide variation points for others to
``bake in quantum characteristics'' to their pet net interpretation,
using the machinery laid out in this paper. The fundamental variation
points are markings, concurrences and direct reachability.
%
%
At higher levels there are further variation points in diagram notation
and net inscriptions.
%

Variation points are a familiar concept in software architecture,
analysis and design, where they serve the separation of the
architectural framework from the more variable architectural
elements. For example in the Model-View-Controller framework, a
reoccurring pattern of software architecture, the aim is to allow
modifications of the view, while minimally, or not at all, affecting
the model and controller part of the software. The same goes for
changes to the data model or the user and software control of the
model. Plugin architectures and feature combinations are further
examples. There are many more variation concepts. All have in common
that parts of the software are parametrised and varied functionally
and as independently as possible.

\subsection{Superpositions of Multisets}
\label{sec:multisets}
We begin with a formal account of multisets which are used for
markings, concurrences and superpositions in QPNs.
Given a set $A$, a multiset over $A$ is a mapping $s: A
\rightarrow\mathbb{N}$. We also write $s\in\multiset{A}$ for brevity
and abbreviate $s(a)$ by $s_a$ ($a\in A$). Let $s,s'\in\multiset{A}$
be two multisets.

The monomial representation of multisets used throughout this paper is
one of many common representations of multisets in mathematics, for
example, in the representation of the prime factorization of an
integer, as a multiset of prime numbers on the one hand, and the
corresponding exponent vector $ev(s)$ indexed in the order of prime
numbers, on the other. In the monomial representation of $s\in
\multiset{A}$, the elements $a\in s$ are raised to their respective
integer multiplicity $s_a$ as their exponent. Multisets are then
formed by multiplication (or concatenation) for brevity. For example
$ab^2c$ is short for $abbc$ and represents the multiset $\{a\mapsto 1,
b\mapsto 2, c\mapsto 1\}$.  A superposition $v$ of multisets is a
function $\multiset{A}\reach{v}K$ mapping multisets $s_i$ to
coefficients $c_i$ in a field $K$, such as the integers $\mathbb{Z}$,
rationals $\mathbb{Q}$, reals $\mathbb{R}$, Gaussian rationals
$\mathbb{Q}[i]$ or complex numbers $\mathbb{C}$. $v$ is also written
as the polynomial $v=c_1s_1+\cdots + c_is_i+\cdots$ with
$c_i=v(s_i)\not=0$, i.e. listing only the non-zero terms.
Multiplication and addition of monomials and polynomials work with the
usual algebraic laws, treating net elements as variable symbols in
multivariate polynomials.

In net theory and the simulation of net behaviour, the vector $ev(s)$ 
of exponents is often used for efficiently representing a corresponding 
multiset $s$. Then multiplication $ss'$ can be equivalently represented 
as $ev(s)+ev(s')$, consistent with $a^{n+m}=a^na^m$.  
We define the partial order $s<s'$ by comparing the exponents 
$s_a<s'_a$ (for all $a\in A$) and $s\leq s'$ iff $s<s'$ or $s=s'$. 
For the monomial as a product, $s\leq s'$ means $s$ divides $s'$ and
can be cancelled in the division $\frac{s'}{s}$. Consequently,
$ev(s')-ev(s)+ev(s'')$, as used for transition firings, shows as
$\frac{s}{s'}s''$ and simplifies to a monomial again. Set operations
extend to multisets as does some algebraic terminology we need on
occasion:
The {\em degree} $deg(s)$ is the largest exponent of $s$, and the
{\em cardinality} (aka {\em total degree}) $|s|$ the sum of the
exponents. We write $s=1$ iff $deg(s)=0$.
$s$ is called a {\em proper subset} of $s'$ iff $s<s'$, and a {\em subset}
iff $s\leq s'$.
{\em Membership} $a\in s$ is defined by $s_a>0$.
The {\em intersection} (aka {\em greatest common divisor}) $s\cap s'$
is the element-wise minimum of the exponents in $s$ and $s'$.
The {\em union} (aka {\em smallest common multiple}) $s\cup s'$ is the
corresponding maximum. 
The {\em set difference} $s\backslash s'$ is defined by element-wise
subtraction of exponents, if the result is non-negative and otherwise
$0$.
If $f$ is a function $A\reach{f}\multiset{B}$ from a set to a
multiset, then $f$ can be lifted naturally to a function
$\multiset{A}\reach{f}\multiset{B}$, s.t., for all
$s\in\multiset{A}$ we have:
$f(s)_b=s_a\times f(a)_b$ for all $a\in A$ and $b\in B$.
This was illustrated in Fig. \ref{fig:concurrence}.

\subsection{Classical Petri Nets}
\label{sec:ptn}
Next, we define Petri nets parametrised at two variation
points. Various extant net classes can then be derived as variations
and restrictions, incl. their quantum variants.
Firstly, a {\em generator} $G$ is a function $Set\reach{G}Set$, such
that $G[P]$ and $G[T]$ are disjoint, when $P$ (the place set) and $T$ 
(the transition set) are.  This captures a range of net extensions.
Secondly, a Boolean function {\em well-formedness}, short \WF, on
$G[N]$, $\multiset{G[N]}$ and $\multiset{G[N]}\times\multiset{G[N]}$
captures constraints appearing in the net literature under different
names, incl. guards and inhibitors. We define $\PGen$ and $\TGen$ as
the maximal subsets of $G[P]$ and $G[T]$ with WF$(\PGen)={\bf true}$
and WF$(\TGen)={\bf true}$, respectively. $G[N]$ is called the set of
{\em generated net elements}. For $m\in\multiset{\PGen}$ and $t\in
\multiset{\TGen}$, we interpret WF$(m,t)$ as $m$ {\em may enable}
$t$. Also we require strictness of WF on multisets and pairs, i.e.,
WF$(m)={\bf false}$ implies WF$(mm')={\bf false}$, WF$(m,m')={\bf
  false}$ and WF$(m',m)={\bf false}$. Informally, a composite cannot
be well-formed, if one of its components is ill-formed. It follows
from strictness, that WF$(1)={\bf true}$.

\begin{definition}[Net structure]
  \label{def:netstructure}
A {\em Petri net structure} $\Net=(P,T,F)$ (short {\em net})
with generator $G$ and well-formedness \WF\ is a structure where $P$
and $T$ are disjoint sets, called {\em places} and {\em
  transitions}. $F$, which is called {\em flow}, is a pair of functions 
$\multiset{\PGen}\rreach{F^-} \TGen\ \reach{F^+} \multiset{\PGen}$.
$N=P\cup T$ is also called the set of {\em net elements} and
$\PGen\cup\TGen$ the set of {\em generated net elements}.
We call $\Fminus(t)$ the {\em input} and
$\Fplus(t)$ the {\em output} of $t$ and the set of elements
$p\in\Fminus(t)$ the {\em preset} of $t$ denoted by
$\preset{t}$. Analogously, the elements of $\Fplus(t)$ form its {\em
  postset} and are denoted $\postset{t}$. We require of $F$ that
$|\preset{t}|+|\postset{t}|>0$.
\end{definition}
For $G[N]=N$ and \WF={\bf true}, we get PTNs. Coloured Petri nets and
algebraic nets pair places with data of some sort in $G[P]$ (see
Sec. \ref{sec:datarich}).
\begin{convention}
  \label{con:netN}
  For the rest of the paper, let $\Net$ be the PN structure
  with the above components.
In the well-known incidence matrix representation of a PTN, $\Fminus$
and $\Fplus$ are represented as $P\times T$-indexed matrices called
the {\em input} and {\em output incidence matrix}, respectively (c.f.,
e.g., \cite{murata1989:PetriNetsProperties,reisig1985:PetriNetsIntroduction}). This generalises to corresponding
$G[P]\times G[T]$-indexed matrices for the net structures above with a
generator $G$.
\end{convention}
\begin{definition}[Enabling and firing]
\label{def:enabling}
Let $m,m'\in\multiset{\PGen}$ and $t\in \multiset{\TGen}$.
Then $m$ is called a {\em marking} and $t$ a {\em concurrence} of
$\Net$.
Now, if $\Fminus(t)\leq m$ and WF(m,t), then $t$ is said to be {\em
  enabled} in $m$. This is abbreviated by $m\reach{t}$.  If
$ev(m')=ev(m)-ev(\Fminus(t))+ev(\Fplus(t))$, then we say $m'$ is {\em
  concurrently (or asynchronously) reachable} from $m$ by {\em firing}
$t$. This is abbreviated by $m\reach{t}_cm'$. We write $m\reach{}_c
m'$ if there is a concurrence $t$, s.t. $m\reach{t}_cm'$. This binary
relation on markings is called {\em direct concurrence
  reachability}. Its transitive closure $\reach{}_c^+$ is called {\em
  concurrence reachability}.  With the restriction $|t|=1$, we
abbreviate $m\reach{t}_cm'$ to $m\reach{t}_1m'$ and $m\reach{}_cm'$ to
$m\reach{}_1m'$. The binary relation $m\reach{}_1m'$ is called {\em
  direct single-transition reachability} and its transitive closure
$\reach{}_1^+$ {\em single-transition reachability}. Moreover, if
$m\reach{t}_c m'$, we call $(m\backslash m',m'\backslash m)$ the {\em
  effect} of $t$. We omit the subscripts $c$ and $1$, if there is no
risk of confusion in the given context.
\end{definition}
Note that the effect of $t$ ignores the intersection of its input and
output.  Importantly, enabling, firing effect and reachability are
defined purely in terms of net structure and WF.  Note that
$\reach{}_c$ is reflexive by definition, given that $1$ is a
concurrence.
%
%
\begin{definition}[Loop-free, reversible, pure and simple] 
\label{def:puresimple}
The net $\Net$ is
  {\em loop-free}, iff $\Fminus(t)\not = \Fplus(t)$ (for all $t\in \TGen$);
  {\em reversible}, iff for all $t\in\TGen$ there is a $t'\in\TGen$, s.t., 
  $\Fminus(t)=\Fplus(t')$ and $\Fplus(t)=\Fminus(t')$;
  {\em pure}, iff $\preset{t}\cap\postset{t}=\emptyset$ (for all $t\in
  \TGen$); and 
  {\em simple}, iff for any $t, t'\in \TGen$: $\Fminus
  (t)=\Fminus (t')$ and $\Fplus (t)=\Fplus (t')$ implies $t=t'$.
\end{definition}
A loop is thus a transition whose input equals its output, making its
effect trivial. Purity means, its inputs and outputs do not share
common factors (intersection) in any multiplicity. Simplicity means,
the input-output pair uniquely defines the transition.  Purity implies
loop-freedom. Simplicity does not exclude impurity or
loop-freedom. For example, the net in Fig. \ref{fig:DoubleSlit} is
simple, but has loops and is impure, while the net in Fig.
\ref{fig:entangled} is simple, pure and loop-free. The definition of
purity and simplicity can also be used to define a respective
equivalence in order to {\em purify} or {\em simplify} the net,
respectively by forming the quotient, after also removing loops in the
case of purification. The resulting net is pure or simple,
respectively, by construction. 

\begin{definition}[System net, reachability set and graph]
\label{def:systemnet}
  $\SysPN=(\Net,M_0)$ is called a {\em system net} with {\em initial
  marking} $M_0\in\multiset{\PGen}$.  The {\em reachability set}
$\reachset_0$ is defined as the set of all markings reachable from
$M_0$ by single-transition reachability. The {\em reachability graph}
$\reachgraph{S}=(V,E)$ is the multigraph\footnote{In a multigraph a
single pair of edges $(u,v)$ may have multiple parallel edges, here
represented as triples $(u,k,v)$. A graph with a single unique edge
for each pair of vertices is called simple in graph theory.} with
  $V=\reachset_0$ and 
  $E=\set{(m,t,m')}{m\reach{t}_1m', t\in \TGen\ {\mathrm and}\ m,m'\in
  \reachset_0}$.  $\SysPN$ is called {\em safe} iff $deg(m)\leq 1$ for
all $m\in\reachset_0$ and {\em reduced} iff for all $t\in\TGen$, there
are markings in $m,m'\in\reachset_0$ with $m\reach{t}m'$.
\end{definition}

It is straightforward to verify that the underlying system net of the
2-qubit QPTN of Fig. \ref{fig:entangled} is loop-free, reversible,
pure, simple, safe and reduced.

\begin{definition}[Equivalent shape and behaviour]
  \label{def:equivalent}
  Let $\SysPN=(\Net,M_0)$ and $\SysPN'=(\Net',M'_0)$ be two system PNs.
  Then $\SysPN$ and $\SysPN'$ are called
  {\em shape-equivalent} under a bijection $f$ between their place
  sets $\PGen\stackrel{f}{\rightleftharpoons}\PGen'$, if $M_0\cong
  M'_0$ and $\reachset_0\cong\reachset'_0$ under the congruence
  uniquely induced by $f$ on $\multiset{\PGen}$. 
  They are called {\em behaviour-equivalent} if they are
  shape-equivalent under $f$ and $f$ moreover extends to a bijection
  $\TGen\stackrel{f}{\rightleftharpoons}\TGen'$, s.t.  their
  reachability graphs are isomorphic under $f$:
  $\reachgraph{S}\cong\reachgraph{S'}$.
\end{definition}
Shape-equivalent system PNs operate on isomorphic markings but
possibly with different transitions and concurrences. With equivalent
behaviour, the nets can simulate each other as their direct
reachability relations mirror each other under the bijection $f$
between their net elements. Behaviour equivalence implies that $f$ is
an isomorphism on their reachability graphs and preserves the direct
concurrence reachability. The reverse is not true, because a
reachability graph isomorphism may not preserve the enabling of
concurrent transitions. For simplicity, henceforth we identify places,
markings and reachability set in two nets of equivalent shape, without
mentioning $f$.

\subsection{Baking Quantum into Petri Net Theory}
\label{sec:baking}

Next, we formalise QPNs and look at the matrix calculus and linear
algebra operators they induce based on their underlying nets.

\begin{definition}[Quantum Petri net]
\label{def:qpn}
  Let $\SysPN=(\Net,M_0)$ be a system PN, then $\QPN=(\SysPN,r)$ is a
  {\em Quantum Petri net} (short QPN) where $r$ is a function
  $\reachset_0\times \multiset{\TGen}\reach{r} \mathbb{C}$. We write
  $r_m(t)$ for $r(m,t)$. A {\em superposition state} (short {\em
    superposition}) of $\QPN$ is any $|\reachset_0|$-dimensional
  complex vector $v$.
\end{definition}

Thus, the underlying net spans an $|\reachset_0|$-dimensional complex
Hilbert space of marking superpositions. The canonical basis vectors
are ${\bf b}_m (m\in\reachset_0)$ with all entries $0$ except a single
$1$ at index $m$. We conveniently abbreviate ${\bf b}_m$ by
$\ket{m}$. Any superposition $v$ can be written as the linear
combination $\sum_m v_m \ket{m}$ of basis vectors. $v$ is also written
as $\ket{v}$ to recall it is a column vector. The inner product of the
Hilbert space is $\braket{u}{v}$, which multiplies the row vector
$\bra{u}$, which is the conjugate transpose of $\ket{u}$, with the
column vector $\ket{v}$. The inner product lends a geometry to this
complex vector space, with the real-valued length
$|u|=\braket{u}{u}$. The distance of two vectors is
$|\ket{u}-\ket{v}|$ and the angle between them is
$\alpha=arccos\frac{\braket{u}{v}}{|u|\times|v|}$.
\begin{definition}[Rate graph]
\label{def:rategraph}
Let $\QPN$ be a QPN as above (Def. \ref{def:qpn}).
The {\em rate graph} of $\QPN$ is the multigraph 
$\rategraph{Q}=(V,E)$ with
  $V=\reachset_0$ and 
$E=\set{(m,t,m')}{m\reach{t}_cm', t\in \multiset{\TGen} {\mathrm
    and}\ m,m'\in \reachset_0}$.  $E_{m,m'}$ denotes the set of edges
between $m$ and $m'$ and we often simply write $m\reach{t}_c m'$
instead of $(m,t,m')\in E_{m,m'}$ given the correspondence between
edges and the direct concurrence reachability.
\end{definition}
Note that the rate graph of a QPN is defined purely in terms of the
underlying system PN, as its concurrence reachability graph.  This
defines the domain of the rate function, in the sense that any direct
concurrence reachability $m\reach{t}_cm'$ has a unique edge in the
rate graph, for which $r_m(t)$ defines the amplitude, and vice versa.
Also note, that $|\reachset_0|=\infty$ is possible in more than one
way. Firstly, the generator $G$ may create infinite nets, i.e.,
$|G[N]|=\infty$. Secondly, even for a very small finite net, the
reachability set may be very large (or even infinite). This generative
power is well-known for nets and used, for instance, in SPNs, where it
is combined with structural (net-level) methods and behavioural
(reachability graph or matrix) methods. However, some numerical
net-theoretic methods for concurrent and stochastic processes require
working directly with the reachability graph and therefore, more often
than not, we ask whether a given system PN is bounded\footnote{A
system PN is called bounded if there is a bound $b\in\mathbb{N}$ with
$deg(m)<b$ for all reachable markings. It is bounded iff its
reachability set is finite.}  and therefore $\reachset_0$ is finite.
This is solved for the classical underlying PN.
We also say a QPN has property $X$ if its underlying system PN has
property $X$. For example $\QPN$ is simple if $\SysPN$ is simple and
shape-equivalent with another QPN if their underlying system PNs are.
%
Any $n$-qubit QPN that juxtaposes $n$ qubits is loop-free, reversible,
pure, simple, safe and reduced (cf. e.g., Fig. \ref{fig:entangled}).
Therefore its rate graph is a simple graph\footnote{ignoring self-loops
associated with $r_m(1)$ edges}. All these are decidable in $\SysPN$.

\begin{definition}[Rate matrix]
\label{def:ratematrix}
Let $\QPN$ be a QPN with $\rategraph{Q}=(V,E)$ and $|V|=n$.  Then its
{\em rate matrix} $\ratematrix{Q}$ is defined as the $n$-dimensional
square matrix, satisfying:
$\ratematrix{Q}[m',m]=\sum_{t,m\reach{t}_cm'} r_m(t)$. The {\em normal
  rate matrix} is defined by normalising the row vectors of
$\ratematrix{Q}$.
%
\end{definition}
 
Let $R$ be a normal rate matrix of a QPN. Then it can be interpreted
on the underlying system PN $\SysPN$ as follows. $Rx=y$ evolves any
superposition $x$ of markings of $\SysPN$ in a single-step
quantum-parallel evolution to a marking superposition $y$ (cf., e.g.,
Equ. \ref{equ:evolution}).
Since every $n^2$ matrix with complex entries is an operator in the
$n$-dimensional complex Hilbert vector space, the following sentence
is a consequence of the above definition.
\begin{corollary}[QPN rate matrices are Hilbert space operators]
\label{cor:ratehilbert}
Let $\QPN=(\SysPN,r)$ be a QPN. Then the rate matrix $\ratematrix{Q}$
and its normal rate matrix are both operators of the complex Hilbert
space $\Hilbert{n}$, with dimension $n=|\reachset_0|$ and state
vectors $a\in\mathbb{A}_0$. If the net is reversible, let
$t^{-1}\in\TGen$ be the reverse transition for every $t\in\TGen$. If
the rate function $r$ is conjugate symmetric, i.e., for every
$m\reach{t}m'$, $r_{m'}(t^{-1})=r_m(t)^*$, then the normal rate
matrix is unitary.
\end{corollary}

\subsubsection{Universality.} Next, we wish to show that QPNs are a universal
computation model in the sense of the quantum circuit model for
quantum computation. Firstly, we show that the operator matrix of any
circuit defines a QPN. Secondly, we represent a specific universal
gate set in terms of QPNs. Thirdly, we demonstrate the compositional
algebraic nature of QPNs for the construction and analysis of hybrid
causal and acausal quantum-parallel processes (cf. proofs in the
appendix).

\begin{theorem}[Universality of QPNs]
\label{theo:universality}
  Any quantum gate circuit defines a QPN $\struct{Q}$ with $R_Q$ the
  operator matrix of the circuit.
\end{theorem}

\begin{theorem}[Clifford+T QPNs]
\label{theo:cliffordT}
 The universal Clifford+T gate set below of 2-qubit and 1-qubit
 circuit matrices has a straightforward representation as QPNs:
 \begin{equation}
  CNOT=
  \begin{bmatrix}
    1 & 0 & 0 & 0 \\
    0 & 1 & 0 & 0 \\
    0 & 0 & 0 & 1 \\
    0 & 0 & 1 & 0
  \end{bmatrix}
  ~~~H=\sqrthalf
  \begin{bmatrix}
    1 & 1 \\
    1 & -1
  \end{bmatrix}
  ~~~S=
  \begin{bmatrix}
    1 & 0 \\
    0 & i
  \end{bmatrix}
  ~~~T=
  \begin{bmatrix}
    1 & 0 \\
    0 & e^{i\pi/4}
  \end{bmatrix}
\end{equation} 
\end{theorem}

For the benefit of interpreting the above universality results, we
briefly recall the circuit model of quantum computation and its notion
of universal gate set, in order to make this paper somewhat
self-contained.  For a detailed treatment, the reader is referred to
\cite{miszczak2012:HighLevelStructuresQuantum}.  Circuit diagrams are made from qubits (lines
oriented from left to right) and logic gates (typically drawn as boxes
or connectors crossing lines vertically). Gates operate on some of the
qubits only (cf. e.g., Fig. \ref{fig:cnoth}). For a gate, the number of
input lines equals that of its outputs.  The function of each logic
gate on its $n$ qubits is a complex $2^n\times 2^n$ matrix on the
corresponding Hilbert subspace $\Hilbert{2^n}$ spanned by the $2^n$
canonical basis vectors, i.e., vectors everywhere $0$ except for a
single position that is $1$.
The operation of two gates $G$ after $F$ is applied graphically in
series from left to right, on the {\em same} $n$ qubit lines. It is
defined by the matrix-matrix multiplication $GF$ of the corresponding
operator matrices.
Top to bottom juxtaposition of two gates (incl. NOP as a special case,
see below) represents parallel composition, defined by the tensor of
the corresponding matrices.
The NOP (no operation) gate is simply represented by continuing the
$n$ qubit lines it operates on, i.e., without showing the NOP box. Its
function is the corresponding $2^n$-squared identity matrix. This
leaves the states of these qubits unchanged. Therefore, NOP can be
inserted were needed, for example in extending a $n$-qubit gate to a
larger number of qubits using the tensor product with the
corresponding identity matrix.

A number of finite gate sets (typically very small sets) have been
identified as universal, i.e., capable of representing any quantum
computation. The Clifford+T gate set above (Th. \ref{theo:cliffordT})
has been proven to have this property.  However, any finite circuit
diagram is equivalent to a finite square operator matrix, when, in
general, continuous-space-time quantum computations may involve
infinite-dimensional Hilbert spaces, i.e., superposition vectors with
an infinite number of positions. Therefore, a gate set is defined as
universal, more subtly, viz.: if an arbitrarily long but finite
sequence of circuits entirely built from this gate set can approximate
any quantum computation to any required precision. While this means
working with limits, it is not an issue for finite-dimensional Hilbert
spaces, which are always {\em complete}, i.e., have limits for any
converging Cauchy sequence, such as those resulting from arbitrarily
long state evolutions from an initial superposition, and converging
ever closer to some limit. Completeness guarantees that such limits
exist as a superposition in the system.

\subsubsection{Hierarchical component architecture.} Petri net theory offers a
rich set of compositional constructions for the causal, parallel and
hierarchical structuring of the underlying system nets of QPNs. For
example, foldings are net morphisms that can partition the place set
$\PGen$ lumping together all elements of a single partition into a
macro-state and consistently re-interpreting markings and transitions,
altogether arriving at a smaller net and generally smaller
reachability graph with lower-dimensional matrices. Beyond net
compositionality, the relational nature of reachability and the
functional character of rates, from single steps to entire QPNs and
their rate graphs, lend linear algebra properties to the resulting
rate matrices, naturally.  However, the linear algebraic
compositionality of the target space is present in the QPNs themselves
already, as the following compositionality theorem shows.
To our knowledge, the generality of this compositionality result is
novel and somewhat surprising, although research in stochastic Petri
nets has used Kronecker algebra
\cite{campos1999:StructuredSolutionAsynchronously,donatelli2001:KroneckerAlgebraStochastic},
however with constraints.
\begin{theorem}[Compositionality of QPNs]
\label{theo:compQPN}
The class of QPNs is closed under the following operations with QPNs
$\QPN=(\SysPN,r)$, $\QPN'=(\SysPN',r')$, and complex numbers
$c,c'\in\mathbb{C}$:
\begin{description}
    \item[~~{\rm zero:}] ${\bf 0}_Q$, is the QPN over any system PN,
      with zero rate function defined as $r_m(t)=0$ for all reachable
      markings $m$ and concurrences $t$ with $m\reach{t}_c
      m'$. $\ratematrix{{0_Q}}$. It follows, that $\ratematrix{Q}$ is
      the all-0 matrix.
    \item[~~{\rm unit:}] ${\bf 1}_Q$, is the QPN over any system PN,
      with the unit rate function defined as $r_m(1)=1$ and $r_m(t)=0~
      (t\not= 1)$ for all reachable markings $m$ and concurrences with
      $m\reach{t}_c m'$. It follows, that $\ratematrix{{1_Q}}$ is
      the identity matrix.
    \item[~~{\rm scaling:}] $c\QPN=(\SysPN,c\times r)$, with $(c\times
      r)_m(t)=c\times r_m(t)$. 
      It follows that $\ratematrix{cQ}=c\times\ratematrix{Q}$.
    \item[~~{\rm product:}] $\QPN\QPN'=(\SysPN'',r'')$ is called the
      monoidal {\em product} (aka concatenation) and defined as
      follows, if $\QPN$ and $\QPN'$ are shape equivalent.
      $\QPN\QPN'$ has the shape of its components. This means the QPNs
      have identical places, markings and reachability set (up to
      isomorphism).
      We require: for all $m,m',m''\in\reachset_0$:
      $E''_{m,m''}=E_{m,m'}\times E'_{m',mm''}$, where
      $E, E'$ and $E''$ are the respective rate graph edge sets.
      I.e., for every pair of concurrence edges $m\reach{t}_c m'$ in
      $E_{m,m'}$ and $m'\reach{t'}_c m''$ in $E'_{m',m''}$ we have the
      contracted edge $m\reach{(t,t')}_c m''$ in $E''_{m,m''}$ and and
      vice versa, with the rate of the contracted edge
      $r''_m(t,t')=r_m(t)\times r_{m'}(t')$.  Finally, we define
      $\TGen''$ as contraction of singleton concurrences $t$ and $t'$
      in the respective component, using the contracted concurrence
      pairs $(t,t')$, $(t,1)$ and $(1,t')$ with appropriate $\Fminus$
      and $\Fplus$ according to the firing sequence $tt'$.
      It follows that
      $\ratematrix{QQ'}=\ratematrix{Q'}\ratematrix{Q}$.
    \item[~~{\rm sum:}] $\QPN+\QPN'=(\SysPN'',r'')$ is called the {\em
      sum} and defined as follows, if $\QPN$ and $\QPN'$ are shape
      equivalent and all $t\in\TGen\cap\TGen'$ satisfy
      $\Fminus(t)=\Minus{F'}(t)$ and $\Fplus(t)=\Plus{F'}(t)$.
      $\QPN+\QPN'$ has the shape of its components. 
      We require $E''_{m,m'}=E_{m,m'}\cup E'_{m,m'}$ for all
      $m,m'\in\reachset_0$, where
      $E, E'$ and $E''$ are the respective rate graph edge sets.
      The rate function $r''$ of the sum is defined s.t.
      $r''_m(t)=r_m(t)$ for $(m,t,m')\in E_{m,m'}\backslash
      E'_{m,m'}$, $r''_m(t)=r'_m(t)$ for $(m,t,m')\in
      E'_{m,m'}\backslash E_{m,m'}$, $r''_m(t)=r_m(t)+r'_m(t)$ for
      $(m,t,m')\in E'_{m,m'}\cap E_{m,m'}$, and $r''_m(t)=0$
      otherwise.  Finally, we define $\TGen''$ as singleton
      concurrences with $m\reach{t}_c m'$ in $E''$ with their uniquely
      defined respective input and output markings.
      It follows that $\ratematrix{Q+Q'}=\ratematrix{Q}+\ratematrix{Q'}$.
    \item[~~{\rm Kronecker product:}] $\QPN\otimes\QPN'$ is the
      disjoint juxtaposition (aka tensor product) of the two QPNs --
      their isolated parallel composition -- in this order.  The
      resulting rate graph satisfies $\rategraph{Q\otimes Q'}=
      \rategraph{Q}\otimes\rategraph{Q'}$ with the usual
      graph-theoretic Kronecker product of graphs. It follows that the
      resulting rate matrix satisfies, $R_{Q\otimes Q'}=R_{Q}\otimes
      R_{Q'}$.
    \item[~~{\rm Kronecker sum:}] $\QPN\oplus\QPN'$ is defined as
      $\QPN\otimes{\bf 1_{Q'}}+{\bf 1_{Q}}\otimes \QPN'$.  It follows
      that the resulting rate matrix $R_{Q\oplus Q'}$ equals the
      Kronecker sum of the component rate matrices
      $R_{Q\oplus Q'}=R_{Q}\oplus R_{Q'}$      
\end{description}
\end{theorem}

The above compositionality lends linearity poperties to quantum Petri
net compositions themselves, including hierarchical composition of
marked cyclic nets, which are among the hallmarks of classical net
architecture. Because stochastic rates are `just' special real-valued
rate functions, this result also offers new forms of compositionality
to SPNs and similar net classes.
In a nutshell, QPN addition is associative and commutative with zero
${\bf 0}$.  While generally non-commutative, monoidal concatenation,
the net equivalent of matrix multiplication, is associative with unit
${\bf 1}$ and distributive over addition. Disjoint juxtaposition, with
the usual interpretation of concurrent transition firing of isolated
subnets, is the free parallel composition. Juxtaposition is
associative, but the order matters for the forward reachability and
the matrix index sets. Juxtaposition results in the Kronecker product
of the rate graphs (graph-theoretically) and of the rate matrices.

As a consequence of the above, we arrive at the following
interpretation.
If we apply the rate matrix $R$ of a QPN to a definite marking
$\ket{m}$, i.e., one of the canonical basis vectors, we obtain the
vector $v=R\ket{m}$ identical to the $m$-th column of $R$. If
$v_{m'}>0$, then $m'$ is reachable from $m$ in the underlying net with
amplitude $v_{m'}$.  If $v_{m'}\not =0$, then $m'$ is either
unreachable in the underlying system from $m$ in a single step
(transition or concurrence). Or else, this underlying reachability
step is rated $0$ by the rate function of the QPN generating $R$.  So,
we can simply read the rated concurrence reachability off the rate
matrix.

An $n$-step evolution of a quantum system can be obtained by
matrix-matrix multiplication. For QPNs the rate matrices work in
similar way.  However rate matrices are forgetful, in that they do
not include the concurrence structure inherent in multisets of places
and transitions.
%

In contrast, the product of QPNs is not forgetful of the concurrency
structure. Each path with non-zero amplitude in the rate graph allows
us to reconstruct a run of the QPN (a kind of PTN occurrence net) of a
length defined by the path and a width defined by the maximal
cardinality of its concurrence steps. The product $QQ'$ reflects the
OR-parallel quantum execution combinatorially joining up steps and
runs through intermediate markings. An $n$-step superposition
evolution of a QPN $Q$ can then be identified with the direct
concurrence reachability in its monoidal power $Q^n$ (of the same
shape as $Q$). Similarly, the QPN sum $Q^{(1\leq n)}:=\sum_{1\leq
  i\leq n}Q^i$ characterises such an evolution of at least $1$ and up
to $n\geq 1$ steps; $Q^0={\bf 1_Q}$ is the identity matrix; and
$Q^{(0\leq n)}:={\bf 1_Q}+Q^{(1\leq n)}$ the evolution of $0$ up to
$n\geq 1$ steps\footnote{Cf. the similarity to products and sums of
adjacency matrices of graphs -- determining the existence of runs of
specified lengths}. Considering the collective OR-parallelism of QPNs
superposition evolutions as superposed runs, we note that these do not
have to follow in lock step. Any mix of possible lengths can be
expressed using sums, products and powers of QPNs, as if unwinding the
(possibly cyclic) QPNs according to their hierarchical composition.

%

It should be noted that the {\em concurrences} underlying the rate
graphs and rate matrices of QPNs, and hence the various
interpretations above {\em are independent of a global time and hence
  independent of a specific observer}. The multiple transitions in a
concurrence may fire entirely asynchronously. But they can also be
entangled in a joint -- rhythmical and resonant -- firing as a
function of the causal net structure, the QPN rates and specific
complex amplitudes in initial superposition.
Because of this, QPNs unify not only classical and quantum computation
but also abstract from various time models associated with the
corresponding net systems, opening them up for various
interpretations, incl. continuous time, discrete time, partial-order
event occurrences, stochastic event occurrences etc. Of course, when
modeling real physical systems rather than abstract quantum
algorithms, any interpretation must be consistent with the quantum
mechanical behaviour, ultimately in terms of very specific Hilbert
spaces and their operator matrices, whether the generator is a QPN or
a quantum circuit diagram.

\section{How To Bake Your Pet Net Class with Quantum Flavour}
\label{sec:variants}
In the literature, many Petri net classes have been defined as
extensions of elementary and of place-transition nets.  Another
approach, taken here, is to parameterise Petri nets and look at
specific actual parameters, variation points and restrictions in the
spirit of software architecture families and product lines
\cite{jazayeri2000:SoftwareArchitectureProduct,bosch2010:IntegrationCompositionImpact,yusuf2013:ParameterisedArchitecturalPatterns}.

For example, we simply define a {\em place-transition net} structure
(short PTN) as a net structure (see Def. \ref{def:netstructure}) with
$G[N]=N$ and WF$={\bf true}$, and, an {\em elementary Petri net}, EPN
for short, as a PTN with $deg(\Fminus(t))=deg(\Fplus(t))=1$ for all
$t\in T$. A QPN with the corresponding restriction is naturally called
Quantum PTN (QPTN) and Quantum EPN (QEPN), respectively.

QPNs that are based, in this way, on our formalisation
(Sec. \ref{sec:formal}) above with specific parameters for the $G$, WF
and other variation points, then also induce a well-defined quantum
interpretation by virtue of their underlying PN variant generating the
relevant reachability relations. Most importantly, the theory (i.e.,
initially the theorems above) is valid and the bridge to quantum
information theory and quantum computation, that we established above,
applies to any of those variants.

We briefly sketch a few variations related to extant Petri net
classes. This demonstration also aims to enable the reader to apply
similar constructions to their own pet Petri net class, without
compromising the quantum interpretation of the QPNs that arise from
them as their pet underlying PN class.

\subsection{Quantum Logical Guards}

Many PN extensions associate guards with transitions. A guard is
typically a Boolean expression inscribed to, or associated with, a
transition.  It may have free variables that may be bound to the
number of tokens in one or more places related to the transition, or
to the token values in coloured nets. Sometimes free variables may be
bound to constants, i.e., user-defined parameters, or to variable
values in an extended notion of marking, e.g., the values of one or
more clocks to represent asynchronous time with several local clocks.
Any free variables can be absorbed in the generator $G$ and result in
transition schemes $t[x_1,\ldots,x_n]$, with concrete instances
$t[v_1,\ldots,v_n]$ for some actual values $v_i$ admissible for the
free variables. WF$(t[v_1,\ldots,v_n])={\bf true}$ then expresses
syntactic well-formedness or semantic well-definedness of such
transition instances in $\TGen$. When a guard is false for a given
well-formed transition instance $t\in\TGen$ and marking $m$, the
transition is disabled even if $\Fminus(t)\leq m$. This is represented
here by WF$(m,t)={\bf false}$. According to Def. \ref{def:enabling}
this implies that the transition is not enabled and hence a
corresponding reachability relation is not present in the reachability
graph. Clearly the remaining definitions and theorems for QPNs remain
well-founded and valid, respectively.

\subsection{Quantum Petri Nets with Inhibitors and Phase Transitions}

\begin{figure}[bht]
\centering
\raisebox{0.05cm}{
  \includegraphics[width=0.45\textwidth]{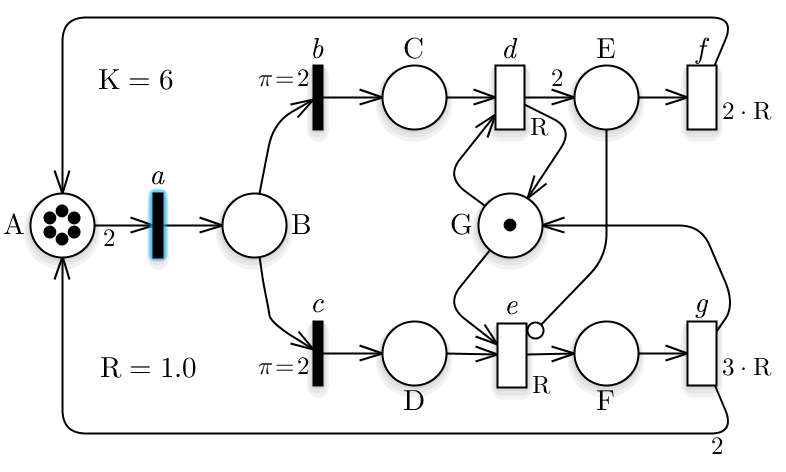}
  }~~
\raisebox{0cm}{
  \includegraphics[width=0.45\textwidth]{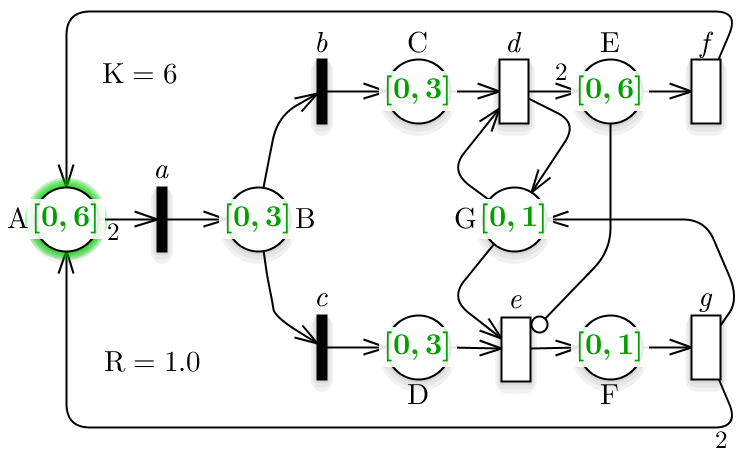}
  }~~
\begin{center}
  \caption{
    \label{fig:gspn}
          {\em GSPNs are a restricted class of QPNs with positive real rates and other restrictions.}
\newline
    {\em Left:} The underlying PTN illustrates inhibition and
    immediate transitions with priorities. Transition $e$ has an
    incoming {\em inhibitor arc} from place $E$, graphically
    represented by a circle head.  {\em Immediate transitions} are indicated by
    solid black bars and have associated {\em priorities} ($\pi>0$) with default $1$.
    {\em Rated transitions} are drawn as open boxes and have lowest priority $0$.
    \newline
        {\em Right:} Analysis of the underlying system PTN on the left
        in the GreatSPN toolkit shows that the system is bounded (with
        bounds shown in places), and (not shown here) that the
        system has two S-invariants (S-semiflow eigenvectors) and two
        T-invariants (T-semiflow eigenvectors) allowing further
        reduction using an eigenvector basis in the matrix
        representation of the underlying net.
  }
\end{center}
\end{figure}

Fig. \ref{fig:gspn} shows a QPN with initial marking of $A^K$, here
with constant $K=6$.  All rates are positive reals and parametrised by
a constant $R$. Considering the restriction of rate functions to
positive real values and the variation of normalisation to a division
by the sum of absolute values (L1-norm) instead of the square root of
the moduli squared of the relevant amplitudes, we arrive at
Generalized SPNs (GSPNs) as a special case of QPNs.

However, there are a few further constructs in GSPNs we need to
consider, such as for instance, the inhibitor arc from place $E$ and
pointing to transition $e$. This disables $e$ when the marking of $E$
is greater or equal to the arc weight, here $1$. For this purpose, the
formal definition of GSPNs includes an inhibition function $H$ similar
in type to $\Fminus$, here modelled as
$\TGen\reach{H}\multiset{\PGen}$.  The inhibition action can thus be
captured in our formalism by WF$(m,t)={\bf false}$ if $H(t)\leq m$.
Immediate transitions, depicted in Fig. \ref{fig:gspn} by solid black
bars, have an explicit priority $\pi$ associated to them in the net,
which defaults to $1$ if not specified. When any immediate transitions
is enabled, only the immediate transitions of highest priority are
enabled.  This can be modeled by WF similar to the above conditions,
given that all these GSPN firing constraints are expressed in terms of
the enabling marking. With successive firings of immediate transitions
and acyclic dependencies (which are required for GSPNs) priority
levels inductively decrease until only {\em timed} transitions, of
default priority $0$, are enabled.  If there is a conflict, a
probability distribution (encoded in the rate restrictions for such
transitions) resolves that said conflict and chooses one of the
conflicting immediate transitions.  Any marking that enables an
immediate transition is considered {\em vanishing} and therefore not
part of the reachability set proper that generates the
reachability. Typically, one constructs a reachability graph with
immediate transitions and then reduces it by contracting all paths
consisting of only vanishing markings. This can be done coherently in
respect of rates by several PN modelling and simulation tools. We omit
the details of the relevant algorithms here and refer to the
literature (cf. e.g., \cite{amparore2016:30YearsGreatSPN}).  The
resulting reduced reachability set can then form the basis for our QPN
theory.

Note that GSPNs with the above rate restrictions and those other
variations are now a special restricted class of QPNs.  In addition,
Quantum GSPNs (QGSPNs) with complex-valued rate functions arise from
the above transliteration of inhibitors, immediate transitions and
priorities.

Historically GSPNs cater to the need of modellers to capture
significant differences in real rates, especially splitting
transitions into those that complete important functions after some
duration and those that are orders of magnitude shorter -- effectively
timeless. Hence immediate transitions are important in this type of
net and likely in QPN applications.  In practice moreover, for
real-time systems modelling, the latter have priorities
associated. Amparore et al. write in
\cite{amparore2016:30YearsGreatSPN}, ``{\em a change of state occurs
  not only because there has been a completion of an activity which
  takes time, but also because there has been a change of some logical
  conditions, which may depend on the current state in a rather
  intricate manner.}''  Physical systems often exhibit rapid phase
transitions and engineered systems rapid mode transitions as a
response to critical sensory input or reaching critical reactivity
levels etc.  Modelling these different types of transitions fully on
the basis of Hilbert operators for QPNs, or CTMCs in the case of
GSPNs, makes their numerical analysis very complex and sometimes
infeasible. Like with real-time systems, we expect that for
real-space-time quantum systems, such structuring mechanism will
empower modellers similarly, to represent the causal and acausal
architecture of quantum systems using both logical and physical
dependencies.

\subsection{Data-Rich Higher-Level Quantum Petri Nets}
\label{sec:datarich}
Many high-level nets have been studied and a variety of net classes
adds data to enrich markings. Among these are Predicate-Transition
nets, various kinds of algebraic nets and Coloured Petri nets.  A
marking in such a net is generally a multiset of data items for each
place.  Different net classes add type-checked data types, tuples and
various other bells and whistles. We capture such extensions by the
generator $G$, which pairs the place set $P$ and data set $D$ and
forms $P\times D$ with projections to the respective place and data
item. Marking $p\in P$ by multiset $a_1^{n_1}\cdots a_n^{k_n}$ in such
a net translates, for our nets, to $(p,a_1)^{n_1}\cdots
(p,a_n)^{k_n}$.  Similar to the above use of WF, the well-formedness
function eliminates ill-formed data pairing with places, or further
restrictions on markings such as out-of-bound multiplicities, which
are common in some of these higher-level nets.

Likewise, transitions are paired with all the data that can reach them
via flow arcs inscribed with data and multiplicity expressions using
variables. We can use transition schemes similar to the encoding of
guards in WF. Thus $G$ generates a data-rich set of net elements,
the flow $F$ between them is generated from the arc inscriptions.
$\Fminus$ and $\Fplus$ remain input and output multisets like before,
for all well-formed and legitimate combinations in the higher-level net.
And finally guards on the transitions are transcribed to WF as already
shown above.

Algebraic specification of partial functions and predicates mimicking
Horn clause logic specification uses a similar construction and has
been applied to nets.  For example, Predicate-Event nets use
many-sorted algebraic specification over nets to generate an ordinary
net structure and its markings modulo theory
\cite{kramer1985:StepwiseConstructionNonsequential,kramer1987:TypesModulesNet,schmidt1991:PrototypingAnalysisNonsequential}
and enriching net interface descriptions with the power of abstract
data types and modules. Analysis methods and executable nets are
implemented there based on term rewriting and a compiled functional
language. In a partial algebra specification, the free generation and
the definedness constraints (here encapsulated\footnote{and leaving
open how this variation point is actualised} in $G$ and WF) use weak
and strong equality in a system of conditional equations.  Weak
equality satisfies: $t\doteq t'$ iff $t$ defined, $t'$ defined and
$t=t'$.  Therefore, we have that $t\doteq t$ iff $t$ is defined. This
makes such an equation also useful in a conclusions of a conditional
equation, to express conditions for definedness. For strong equality
$\equiv$ we have that: $t\equiv t'$ iff ($t$ defined or $t'$ defined)
implies $t\doteq t'$. That is to say, if one side of the equation is
defined, the other must be defined and the two must be equal.

\section{Related and Future Work}
\label{sec:related}
%
Ojala’s group \cite{ojala2004:ModelingAnalysisMargolus} used Coloured Petri nets for
the analysis of a certain class of quantum cellular automata designs,
by example. They achieve a considerable state space reduction compared
to the cellular automata. They do not make full use of the
asynchronous nature of Petri nets, nor arrive at the kind of
universality that characterises QPNs. They model the control structure
of the cellular automaton explicitly as a classical PN with complex
values as token colours and do not aim at a universal representation
of quantum computations.  Their encoding achieves a considerable
reduction in the state space compared to that of the equivalent
quantum cellular automaton.  This work was one of the motivations for
our approach, in the hope that causal modelling, immediate transitions
and other well-known constructions from classical Petri nets can
provide a hybrid classical-quantum design with such gains despite
offering universality and quantum-only QPNs in the pure quantum case.

Much work exists on classical hybrid and fluid nets, in which causal
structuring and stochasticity is mixed with real and integer markings
of places
\cite{ciardo1989:SPNPStochasticPetri,horton1998:FluidStochasticPetri}. Especially
their connection with GSPNs has helped us envisage a general approach
to superposition and entanglement with its difference to classical
stochasticity.  For reasons of brevity, we have omitted detailing
hybrid quantum nets, in this paper.  Suffice it to mention, that one
can always partition the set of places into a finite number of place
kinds carrying different kinds of tokens, including Boolean, integer,
complex integer ($\mathbb{Z}[i]$), real or complex values.  Then
transitions are classified to form actions on these, including on
mixtures of these kinds. This path has been well trodden in Petri net
theory and a number of connections can be drawn to QPNs, although, to
our knowledge, quantum has never before been combined directly with
these nets.


Future work may be fruitful in
\begin{enumerate}
\item {\em graphical calculi} combining QPNs with circuit diagrams
  such as the ZX calculus
  \cite{vandewetering2020:ZxcalculusWorkingQuantum}.  There are
  well-known theorems relating partial-order semantics of Petri nets
  with hierarchical message sequence charts and rational algebraic
  theories of partially ordered event structures, which may benefit
  quantum calculi.  Dually, the circuit diagrams of QPNs may lend
  themselves for the execution of QPNs on real quantum computers.
  QPNs and perhaps nets more generally may benefit from the complexity
  and execution time advantages of quantum computing.
\item {\em machine learning and optimisation} of quantum protocols
  leveraging both the quantum computing speed advantage and advances
  in classical hardware for machine learning has received much
  attention recently. Novel approaches such as quantum and classical
  co-design \cite{jiang2021:CodesignFrameworkNeural} may be especially
  applicable to our approach so fundamentally intertwining the two
  aspects.
\item {\em asynchronous quantum automata} have not received much
  attention. Petri nets have been studied for decades as
  asynchronously generating and accepting automata for formal
  languages. A rich set of theorems exists on their decidability and
  expressive power in comparison to other automata and formal language
  approaches. A corresponding analysis of quantum automata is still in
  its infancy. QPNs may assist revisiting these from the perspective
  of the underlying PNs and their composition.
\item {\em open quantum interaction protocols} at the interface of
  classical and quantum components of QPN systems. Much current focus
  in quantum hardware is naturally on closed quantum
  systems. Measurements are typically final, collapse the system as a
  whole and require the restart of the entire system.  With their
  hybrid causal and acausal structure, QPNs enable a true hybrid
  between open and closed systems, where the network architecture can
  increase coherence from the collapse due to partial measurement.
\end{enumerate}

\section{Conclusion}
\label{sec:conclusion}

This paper introduced a novel diagrammatic model for quantum
information processing, Quantum Petri nets (QPNs). It adds to the rich
theory of Petri nets new results that are far from straightforward,
including the universality of QPNs for quantum computation and their
architectural compositionality that reaches from the well-known
classical net structure to the operator matrices of Hilbert vector
spaces. To our knowledge, these results are {\em novel and original}.
While compositionality is expected for the rate matrices generated by
these nets, the matrices are forgetful of the net structure. However,
the compositionality we explore here is already inherent in the net
structure of the highly parallel quantum processes. QPNs therefore
lend themselves for a formal component-based software architecture
with well-defined entanglement, teleportation and tunnelling across
component boundaries beside the traditional interface protocols, for
which nets have been studied in theory and practice.
QPNs and their theory are based on the formalism of Petri nets which
dates back well before the notion of quantum computation was
formulated, to Petri's 1962 PhD thesis, which itself was developed
with reference to principles of relativity and quantum uncertainty
\cite{schmidt2019:PetriNetsNext}. Over the many decades since, nets
have become a widely used and standardized notation for concurrent
processes in parallel and distributed software modelling and other
process-rich domains with several directly and indirectly associated
ISO and DIN standards.
QPNs are reconnecting with Petri's original motivation for his nets.
Abramsky wrote in \cite{abramsky2006:WhatAreFundamental}
\begin{quote}
Petri’s thinking was explicitly influenced by physics. To a large
extent, and by design, net theory can be seen as a kind of discrete
physics: lines are time-like causal flows, cuts are space-like
regions, process unfoldings of a marked net are like the solution
trajectories of a differential equation. This acquires new
significance today, when the consequences of the idea that information
is physical are being explored in the rapidly developing field of
quantum informatics.
\end{quote}

The paper demonstrated that a separation of concerns can be achieved
between the classical concurrency structures, typical for Petri nets,
and the specific quantum character of entanglement, teleportation and
tunnelling. This separation allows modellers to apply Petri net
methods and tools to QPNs and leverage them for quantum information
processing. Moreover, likely novel results in either field may
accelerate advances in the other through a joint focus on the
orthogonal connection that QPNs show is possible for the two fields of
concurrent and quantum information processes. To this end, this paper
included a number of related and future research problems.

Modern {\em systems architecture requires a dialog} between hardware
and software platform designers, compiler writers, software library
engineers and application software developers, whether for a highly
integrated multi-core tablet and single-user workstation, or a
high-performance supercomputer. The same will undoubtedly be required
of future hybrid quantum and classical systems architectures for
networked distributed quantum systems accessible via cloud services
and platforms. Such services architectures are currently nascent in
commercial offerings. Petri nets have served this dialog in classical
distributed systems as a visual user-friendly and at the same time
mathematically strong tool alongside other strong representations as a
{\em lingua franca crossing fields of expertise}. As part of the
formal treatment of quantum processes in the framework of net theory,
the paper therefore identified architectural variation points in its
formal and informal constructions. We used insights from our prior
research in software architecture design and verification.  We
sketched how the resulting architectural variability can be applied
from elementary Petri nets to Generalized Stochastic Petri nets with
or without colours. A variety of QPN models, their compositionality
and architectures may then be utilised across several classes of Petri
nets, contrasted with, and applied to, real quantum software services
based on the circuit model of quantum computation.

The paper started by a gentle introduction to quantum computation with
QPNs to appeal to the `rest of us': software engineers, practitioners
and computer scientists less familiar with the technical details of
quantum mechanics and their vector spaces than with diagrammatic
models for software programs and their computational processes, in
particular state-machine based concurrent processes such as espoused
in Petri nets and UML architecture diagrams and dynamic models. The
aim was not only to introduce QPNs, but also to recognise and
demonstrate -- before diving into the requisite formalisation -- that
a {\em quantum software engineering narrative is needed and possible},
with minimal knowledge of complex numbers, some basic familiarity with
high-school algebra, and almost no knowledge of vector spaces, at
least at the introductory level. The hope is that this diagrammatic
approach, or perhaps its combination with other suitable and familiar
software models and programming language constructs, may provide a
more gradual entrance ramp to the highway of quantum computing, that
is bound to become a fast multi-lane freeway. Current on-ramps are
steep and access is kept limited to the 'privileged' through a mix of
physics, applied mathematics and theoretical computer science
terminology and theory, yet to be harmonised and standardised,
requiring a steep learning curve, and partly mired in unnecessary
complexity and, at least in popular science, in myth and hype. Through
the architectural compositionality results for QPNs, one might hope,
that hybrid classical and quantum software design can be based on a
diagram-plus-program approach, with verification in graphs and nets,
yet ease of design following principles of modularity, information
hiding and separation of concerns. These hallmarks of software
engineering may empower domain experts and a future open-source
quantum software development community to leverage both classical
methods and advances in quantum computing.


\bibliographystyle{splncs04}
\bibliography{hwszotero20210326}

\ifdefined\DeclarePrefChars\DeclarePrefChars{'’-}\else\fi
\begin{thebibliography}{10}
\providecommand{\url}[1]{\texttt{#1}}
\providecommand{\urlprefix}{URL }
\providecommand{\doi}[1]{https://doi.org/#1}

\bibitem{abramsky2006:WhatAreFundamental}
Abramsky, S.: What are the {{Fundamental Structures}} of {{Concurrency}}?
  Electronic Notes in Theoretical Computer Science  \textbf{162},  37--41
  (2006). \doi{10.1016/j.entcs.2005.12.075}

\bibitem{amparore2018:EfficientModelChecking}
Amparore, E., Donatelli, S.: Efficient model checking of the stochastic logic
  {{CSLTA}}. Performance evalution  \textbf{123-124},  1--34 (2018).
  \doi{10.1016/j.peva.2018.03.002}

\bibitem{amparore2016:30YearsGreatSPN}
Amparore, E.G., Balbo, G., Beccuti, M., Donatelli, S., Franceschinis, G.: 30
  years of {{GreatSPN}}. In: Fiondella, L., Puliafito, A. (eds.) Principles of
  Performance and Reliability Modeling and Evaluation: {{Essays}} in Honor of
  Kishor Trivedi on His 70th Birthday, pp. 227--254. {Springer International
  Publishing} (2016). \doi{10.1007/978-3-319-30599-8\_9}

\bibitem{berthomieu2018:PetriNetReductions}
Berthomieu, B., Le~Botlan, D., Dal~Zilio, S.: Petri net reductions for counting
  markings. In: Gallardo, M.d.M., Merino, P. (eds.) Model Checking Software.
  pp. 65--84. {Springer} (2018)

\bibitem{bosch2010:IntegrationCompositionImpact}
Bosch, J., Bosch-Sijtsema, P.: From integration to composition: {{On}} the
  impact of software product lines, global development and ecosystems. Journal
  of Systems and Software  \textbf{83}(1),  67--76 (2010).
  \doi{10.1016/j.jss.2009.06.051}

\bibitem{campos1999:StructuredSolutionAsynchronously}
Campos, J., Donatelli, S., Silva, M.: Structured solution of asynchronously
  communicating stochastic modules. IEEE Transactions on Software Engineering
  \textbf{25}(2),  147--165 (1999). \doi{10.1109/32.761442}

\bibitem{castelvecchi2017:IBMQuantumCloud}
Castelvecchi, D.: {{IBM}}'s quantum cloud computer goes commercial. Nature
  (London)  \textbf{543}(7644),  159--159 (2017)

\bibitem{ciardo1989:SPNPStochasticPetri}
Ciardo, G., Muppala, J.K., Trivedi, K.S.: {{SPNP}}: {{Stochastic Petri Net
  Package}}. In: {{PNPM}}. vol.~89, pp. 142--151 (1989)

\bibitem{desel2015:ConceptsPetriNets}
Desel, J., Reisig, W.: The concepts of petri nets. Software and systems
  modeling  \textbf{14}(2),  669--683 (2015)

\bibitem{donatelli2001:KroneckerAlgebraStochastic}
Donatelli, S.: Kronecker {{Algebra}} and ({{Stochastic}}) petri nets: {{Is}} it
  worth the effort? In: Lecture Notes in Computer Science. pp. 1--18.
  {Springer} (2001). \doi{10.1007/3-540-45740-2\_1}

\bibitem{heiner2013:MARCIEModelChecking}
Heiner, M., Rohr, C., Schwarick, M.: {{MARCIE}} – model checking and
  reachability analysis done efficiently. In: Application and Theory of Petri
  Nets and Concurrency, pp. 389--399. Lecture Notes in Computer Science,
  {Springer Berlin Heidelberg} (2013)

\bibitem{horton1998:FluidStochasticPetri}
Horton, G., Kulkarni, V.G., Nicol, D.M., Trivedi, K.S.: Fluid stochastic
  {{Petri}} nets: {{Theory}}, applications, and solution techniques. European
  Journal of Operational Research  \textbf{105}(1),  184--201 (1998).
  \doi{10.1016/S0377-2217(97)00028-3}

\bibitem{jazayeri2000:SoftwareArchitectureProduct}
Jazayeri, M., Ran, A., van~der Linden, F.: Software Architecture for Product
  Families: {{Principles}} and Practice. {Ad­di­son-Wes­ley, Reading, MA,
  USA} (2000)

\bibitem{jiang2021:CodesignFrameworkNeural}
Jiang, W., Xiong, J., Shi, Y.: A co-design framework of neural networks and
  quantum circuits towards quantum advantage. Nature Communications
  \textbf{12}(1),  669--683 (2021)

\bibitem{kimble2008:QuantumInternet}
Kimble, H.J.: The quantum internet. Nature (London)  \textbf{453}(7198),
  1023--1030 (2008)

\bibitem{kramer1985:StepwiseConstructionNonsequential}
Krämer, B.: Stepwise construction of non-sequential software systems using a
  net-based specification language. In: Rozenberg, G. (ed.) Advances in Petri
  Nets 1984. pp. 307--330. {Springer} (1985). \doi{10.1007/3-540-15204-0\_18}

\bibitem{kramer1987:TypesModulesNet}
Krämer, B., Schmidt, H.W.: Types and modules for net specifications. In:
  Concurrency and Nets, pp. 269--286. {Springer} (1987)

\bibitem{miszczak2012:HighLevelStructuresQuantum}
Miszczak, J.: High-{{Level Structures}} for {{Quantum Computing}} : {{High
  Level Structures}} for {{Quantum Computing}}. {Morgan \& Claypool Publishers}
  (2012)

\bibitem{murata1989:PetriNetsProperties}
Murata, T.: Petri nets: {{Properties}}, analysis and applications. Proceedings
  of the IEEE  \textbf{77}(4),  541--80 (1989)

\bibitem{ojala2004:ModelingAnalysisMargolus}
Ojala, L., Penttinen, O.M., Parviainen, E.: Modeling and {{Analysis}} of
  {{Margolus Quantum Cellular Automata Using Net}}-{{Theoretical Methods}}. In:
  Cortadella, J., Reisig, W. (eds.) Applications and {{Theory}} of {{Petri
  Nets}} 2004: 25th {{International Conference}}, {{ICATPN}} 2004, {{Bologna}},
  {{Italy}}, {{June}} 21–25, 2004. {{Proceedings}}, pp. 331--350. {Springer
  Berlin Heidelberg} (2004). \doi{10.1007/978-3-540-27793-4\_19}

\bibitem{petri1980:IntroductionGeneralNet}
Petri, C.A.: Introduction to general net theory. In: Brauer, W. (ed.) Net
  Theory and Applications : {{Proceedings}} of the Advanced Course on General
  Net Theory, Processes and Systems ({{Hamburg}}, 1979). Lecture Notes in
  Computer Science, vol.~84, pp. 1--20. {Springer-Verlag, Berlin} (1980)

\bibitem{popkin2017:ChinaQuantumSatellite}
Popkin, G.: China's quantum satellite achieves spooky action at a distance. Sci
  Mag  (2017). \doi{10.1126/science.aan6972}

\bibitem{reisig1985:PetriNetsIntroduction}
Reisig, W.: Petri Nets: {{An}} Introduction. {{EATCS}} Monographs on
  Theoretical Computer Science, {Springer-Verlag, Berlin, Germany} (1985)

\bibitem{schmidt1991:PrototypingAnalysisNonsequential}
Schmidt, H.W.: Prototyping and analysis of non-sequential systems using
  predicate-event nets. Journal of Systems and Software  \textbf{15}(1),
  43--62 (1991). \doi{10.1016/0164-1212(91)90076-I}

\bibitem{schmidt2019:PetriNetsNext}
Schmidt, H.W.: Petri nets: {{The}} next 50 {{Years}}—{{An}} invitation and
  interpretative translation. In: Reisig, W., Rozenberg, G. (eds.) Carl Adam
  Petri: {{Ideas}}, Personality, Impact, pp. 45--66. {Springer International
  Publishing} (2019). \doi{10.1007/978-3-319-96154-5\_7}

\bibitem{steffen2011:QuantumComputingIBM}
Steffen, M., DiVincenzo, D.P., Chow, J.M., Theis, T.N., Ketchen, M.B.: Quantum
  computing: {{An IBM}} perspective. IBM Journal of Research and Development
  \textbf{55}(5),  13:1--13:11 (2011). \doi{10.1147/JRD.2011.2165678}

\bibitem{vandewetering2020:ZxcalculusWorkingQuantum}
van~de Wetering, J.: Zx-calculus for the working quantum computer scientist.
  {Online}, {Cornell University, USA} (2020),
  \url{https://arxiv.org/abs/2012.13966}

\bibitem{yanofsky2008:QuantumComputingComputer}
Yanofsky, N.S., Mannucci, M.A.: Quantum {{Computing}} for {{Computer
  Scientists}}. {Cambridge University Press} (2008)

\bibitem{yusuf2013:ParameterisedArchitecturalPatterns}
Yusuf, I., Schmidt, H.: Parameterised architectural patterns for providing
  cloud service fault tolerance with accurate costings. In: Proc. of
  {{CBSE}}'13. pp. 121--130. {ACM} (2013). \doi{10.1145/2465449.2465467}

\end{thebibliography}

\section*{Appendix}

The following technical notations and proofs are provided as
supplementary material in the {\tt arXiv} version of the submitted
paper, to be filed, and in supplementary material at Springer,
should the paper be accepted.  They are included here for
self-containedness of the submitted paper during the review
process. The supplementary material will possibly include code for
simulating small QPNs.

\subsection{Hilbert Space Notation}
Our notation for a Hilbert space ${\bf H}$ deserves some comments, as
there are different variants for representing inner products and
operator matrices, with their usual properties.
%
We use the Dirac notation, i.e., the so-called `bra-ket' notation
$\braket{u}{v}=u^\dagger\cdot v$ to denote the inner product of two
vectors $u$ and $v$, where $\dagger$ is the conjugate transpose.  The
inner product produces a complex number by multiplying a row vector
with a column vector. This is useful for a comparison of vectors, for
example, to project a vector in the direction of another, to check
orthogonality of two vectors and other measurement related
calculations. In particular, given two superpositions $\ket{u}$ and
$\ket{v}$, the complex amplitude $c$ of a Hilbert space state
transition $\ket{u}\reach{c}\ket{v}$ can be computed by
$c=\braket{u}{v}$.

The use of $\ket{u}$ stresses that a Hilbert space vector is a column
vector $\ket{u}=[u_1,\ldots,u_n]^T$ and avoids the cumbersome, and in
the context of nets ambiguously overloaded, use of $T$ for
transposition. $\ket{u}$ is pronounced `ket-u'.
%
Any object that uniquely identifies a Hilbert space vector according
to some convention in the given context can be placed inside the ket
brackets.  For example, a marking $m$ identifies the canonical basis
vector $\ket{m}=[0,\ldots,1,\ldots,0]^T$ that is $0$ everywhere,
except for a $1$ in the position indexed by $m$. Similarly
$\ket{c_1m_1+\cdots+c_nm_n}= c_1\ket{m_1}+\cdots+c_n\ket{m_n}$
represents the column vector that is the linear combination of marking
basis vectors spanning the Hilbert space over a system Petri net.

The corresponding left part $\bra{u}$ of a bra-ket is pronounced
`bra-u'. On the one hand $\bra{u}$ represents a conjugate transpose,
with $\bra{u}=\ket{u}^\dagger$ and
$\ket{u}=\bra{u}^\dagger$. $\bra{u}$ can therefore be written as the
row vector $[u_1^*,\ldots,u_n^*]$.  On the other hand, not being a
column vector, a bra is not a state vector in ${\bf H}$ but represents
the linear function ${\bf H}\reach{\bra{u}}\mathbb{C}$,
s.t. $\bra{u}(v)=\braket{u}{v}=\sum_iu_{i}^{*}\times v_i$.

%
The following are the basic properties of the above inner product in
bra-ket notation, with other well-known properties implied by them.
\begin{enumerate}
  \item
  conjugate symmetric: $\braket{x}{y}=\braket{y}{x}^*$;
  \item
  positive definite: $\braket{x}{x}>0$ iff $x\not=0$, and $\braket{x}{x}=0$ iff $x=0$; 
  \item
  right linear: $\braket{x}{ay+bz}=a\braket{x}{y}+b\braket{x}{z}$;
  \item
  left antilinear:
  $\braket{ax+by}{z}=a^*\braket{x}{z}+b^*\braket{y}{z}$ (follows from
  2 and 3 above).
\end{enumerate}
In particular, we have made use of linearity of a complex Hilbert
space, when moving scalars in and out of superposition sums, and
generally of related algebraic properties of sums and products of
kets.
%
Note that these properties
also apply to infinite-dimensional Hilbert spaces, then with integrals
instead of sums. Hence the notation applies to unbounded nets.

For a tensor product of, say, three qubits $Q\otimes Q\otimes Q$,
conventionally, one also abbreviates a composite (tensor) state to
$\ket{q_1}\ket{q_2}\ket{q_3}$, where a $q_i$ represents the state of
the $i$-th qubit, or even shorter to
$\ket{q_1q_2q_3}=\ket{q_1}\ket{q_2}\ket{q_3}$. For example, drawing on
such conventions from mathematical physics, we have used $\ket{00}$ to
represent the basis vector of the initial marking $00$ of two
juxtaposed qubit nets. Taking objects into the kets is versatile as it
avoids the exponential combinatorics if used properly. But this
convention requires caution, to avoid ambiguity. It also requires
careful manipulation in sums and products.  In particular recall that,
by definition, an entangled superposition is not expressible as a
(tensor) product of subsystem states.

\subsection{Proofs}

\begin{proof}[Th. \ref{theo:universality}] 
Let $C$ be a quantum circuit diagram over $n$ qubits and $A$ its
$2^n\times 2^n$ complex operator matrix. $A$ represents an operator in
the complex Hilbert space $\Hilbert{2^n}$.  By
Cor. \ref{cor:ratehilbert}, QPNs are equivalent to complex Hilbert
space operators.  Therefore, to prove this theorem, we wish to
construct a QPN Q operating on $\Hilbert{2^n}$ with
$\ratematrix{Q}=A$. Constructing this QPN constitutes a direct proof
of Th. \ref{theo:universality}.

The corresponding QPTN $\QPN$ with underlying system PTN $\SysPN$
juxtaposes $n$ qubits $P_1,\ldots P_n$ analogously to the $2$ qubits
of Fig. \ref{fig:entangled}. The initial marking of this QPN is the
combination of the initial markings of the component nets, i.e.,
$M_0=P_{1,0}\ldots P_{n,0}$ or short $0\cdots 0$ ($n$ times) by our
convention for markings of juxtaposed nets. It is straightforward to
show that $\SysPN$ is loop-free, reversible, pure, simple, safe and
reduced. These are all properties of the underlying system
PN\footnote{and easily verifiable using standard PN model checkers.}.
Safe markings in $\SysPN$ are thus bit strings of length $n$. Because
of purity and simplicity, the single-transition reachability graph
$\reachgraph{S}$ has a unique edge $m\reach{u_i}_1m'$ for all $m$ with
$m_i=0$ and $m'_i=1$, where $m'$ is identical to $m$ in all bit
positions other than $i$.  Accordingly, there is a unique reverse edge
$m'\reach{d_i}_1m$ in the other direction. This graph is therefore
simple.
Consequently, every conceivable safe marking ($deg(m)\leq 1$) is
reachable from the initial state, and any such marking is reachable
from any other in a finite sequence of single-transition steps.
Consequently there are $2^n$ markings, spanning an $2^n$-dimensional
complex vector space with the inner product and other basic operations
above. Therefore this vector space is the complex Hilbert space
$\Hilbert{2^n}$. The initial marking $0\cdots 0$ ($n$ times) is the
basis vector $\ket{0\cdots 0}=\ket{0}\cdots\ket{0}$ ($n$ times).
Any pair of {\em different} markings $m$ and $m'$ has a unique
concurrence $t$ with $m\reach{t}_cm'$ (in $\SysPN$) defined by the bit
positions, in which $m$ and $m'$ differ. $t$ is either a single
transition, or a concurrence of more than one transition. Therefore,
$\rategraph{Q}$ is a simple graph, too, with the exception of loops
for unit concurrences $m\reach{1}_cm$. (Recall that standard graph
theory defines simple directed multi-graphs as graphs that are
loop-free and have edge multiplicity $1$, i.e., unique directed edges,
for all pairs of vertices.)

We now read the rate function $r$ for $\QPN$ off the given circuit
matrix $A$: $r_m(t)=A_{m',m}$ for $m\reach{t}_cm'$.

It remains to show that $\ratematrix{Q}=A$. However, this is trivial,
because above we already showed that $\rategraph{Q}$ is simple, except
for the unit concurrence self-loops. And the rates associated to these
unique edges are exactly those in the diagonal of $A$.  Hence the sum
in Def. \ref{def:qpn} has a single term, viz., $A_{m,m'}$.  \qed
\end{proof}

The proof above relies on (a) the interpretation of a complex circuit
diagram as a single uniquely defined matrix and (b)
universality results for circuits, proven elsewhere.  Because of the
flattening of the circuit into its matrix, we call the above QPN
$\QPN$ the {\em flattened circuit} QPN. The flattened circuit QPN
ignores the architecture of the circuit and is only of interest
mathematically or in terms of low-level translations.  Theorems
\ref{theo:cliffordT} and \ref{theo:compQPN} are more component-based.

\begin{proof}[Th. \ref{theo:cliffordT}]
We have already encountered the QPNs for the CNOT and Hadamard gates
in Fig. \ref{fig:cnoth}.  Here we only need to prove that the
remaining gates of the universal Clifford+T gate set have
corresponding QPNs. This is a straightforward consequence of Theorem
\ref{theo:universality}. Consequently, QPNs can represent the
Clifford+T universal gate set.\qed
\end{proof}

We note that standard circuit models include several other commonly
used gates, which can all be represented as relatively compact
circuits in the Clifford+T gate set and therefore, as QPNs.

The next proof and its Th. \ref{theo:compQPN} guarantee that the
composition of QPN nets themselves can capture the stepwise and
hierarchical composition of qubit circuits, as a special case of
'net-plus-transition-rates' network architecture, describing an
overall system by its components, their interdependency via interfaces
and their interaction with an environment. For the QPNs this includes
classical causal structure because the underlying net is a classical
net and rates acausally constraining the independent firing of
transitions in concurrences as well as the causal firing. Therefore,
this QPN compositionality -- applied to just qubit circuit diagrams --
does not only cover the compositional structure of the dozen or so of
standard gates (via their representation of the equivalent Clifford+T
QPNs above) and that of the more complex quantum circuits expressed in
several tens of interesting quantum protocols in existence today and
characterising the quantum advantage over classical protocols and
processes. It also covers the compositionality of hybrid causal and
acausal control in terms of system PNs and rate functions in QPNs.

\begin{proof}[Th. \ref{theo:compQPN}] For each of the points, we have to prove
  firstly, that the corresponding construction results in a
  well-defined QPN. Secondly, we need to prove the given rate matrix
  equation for the resulting QPN. The algebraic properties of the
  different operations on QPNs follow from those on the corresponding
  matrices, which forgetfully summarise one-step reachability via one
  or more parallel edges between two different markings. Recall that
  the diagonal of the rate matrix is directly defined by $r_m(1)$ for
  all the markings. It is independent of the (single-transition)
  reachability graph $\reachgraph{S}$ of the underlying system and
  reflected in unique self-loop edges $m\reach{1}_cm$ in the rate
  graph $\rategraph{Q}$ of the QPN.

  The proof is straightforward for ${\bf 0}$, ${\bf 1}$ and scaling,
  as these operations are defined purely in terms of the rate function
  component of a QPN without affecting the underlying net. Therefore,
  we omit these here, for brevity.

  The product $\QPN''=\QPN\QPN'$ constructs $\QPN''=(\SysPN'',r'')$
  with underlying system PN $\SysPN''=(\Net'',M_0)$. $\SysPN''$ shares
  the place set $\PGen$, place multisets, initial markings and
  reachability set $\reachset_0$ with the component QPNs, up to
  isomorphism. For brevity, we treat the corresponding set as
  equal. The same goes for other compositions, in which shape
  equivalence of the components is required.

  The transitions and concurrences of $\SysPN''$ are obtained by
  contraction of a pair $(t,t')$ with $t$ from $\QPN$ and $t'$ from
  $\QPN'$. Its transition set is defined by single concurrence
  reachability relations and thus reduced to transition pairs $(t,t')$
  where both $t$ and $t'$ can fire in each their component QPNs. The
  concurrence relation defined for $\QPN''$ is consistent with the
  flow definition for single transitions. Likewise the rate function
  is well-defined.  Hence $\SysPN''$ is well-defined.  Note however,
  that the net flow and rate functions in the component nets may
  differ. Hence their rate matrices differ in general.

  Now for a fixed pair of markings $m,m''\in\reachset_0$ (in the shape
  common to the three QPNs) we have that:
  
  $\ratematrix{QQ'}[m'',m]=\ratematrix{Q''}[m'',m]=\sum_{t'',m\reach{t''}m''} r''_{m}(t'')$ (by rate matrix Def. \ref{def:ratematrix})

  $=\sum_{t,t',m',m\reach{t}m',m'\reach{t'}m''} r_m(t)\times r'_{m'}(t')$ (by def. of product with $t''=(t,t')$)
  
  $=\sum_{m'\in\reachset_0}(\sum_{t,m\reach{t}m'}r_m(t))\times (\sum_{t',m'\reach{t'}m''}r'_{m'}(t'))$ (index rearrangement and
  distributivity)
  
  $=\sum_{m'\in\reachset_0}(\sum_{t',m'\reach{t'}m''}r'_{m'}(t'))\times(\sum_{t,m\reach{t}m'}r_m(t))$ (commutativity of scalar multiplication)

$=\sum_{m'\in\reachset_0}\ratematrix{Q'}[m'',m']\ratematrix{Q}[m',m]$ (substitute rate matrix Def. \ref{def:ratematrix} for $Q$ and $
Q'$)

  $=(\ratematrix{Q'}\ratematrix{Q})[m'',m]$ (def. matrix multiplication)

  Hence $\ratematrix{Q''}=\ratematrix{QQ'}=\ratematrix{Q'}\ratematrix{Q}$.

  For the sum, similarly, the well-definedness is
  straightforward. Notice that shared transitions of the two
  components are required to agree in $\Fminus$ and $\Fplus$.  This
  means the intersection of the corresponding rate graph edge sets is
  well-defined in terms of net structure and reachability.  On this
  intersection, for every pair of reachable markings $m$ and $m'$, the
  rate function computes the sum of the component rate
  functions. Edges in $\QPN'$ outside of the intersection take their
  rates from the respective component.  This means the sum of QPNs is
  well-defined.

  It is also straightforward to see that the rate matrix sum
  $R''[m',m]=\sum_{t'',m\reach{t''}m'}
  r''_{m}(t'')=\sum_{t,m\reach{t}m'}r_m(t)+\sum_{t',m\reach{t}m'}r'(t')$
  contains all the terms from both matrices, because of the sums used
  in the intersection of edges sets $E_{m,m'}\cap E'_{m,m'}$.

  Hence $\ratematrix{Q+Q'}=\ratematrix{Q}+\ratematrix{Q'}$.
 
  Next we look at the Kronecker algebra composition operations
  $\otimes$ and $\oplus$. The tensor product is an abstract and
  general construction and applied in many branches of mathematics.
  For matrices, the tensor product boils down to the Kronecker product
  $\otimes$ below, one of the operations of Kronecker algebras. The
  Kronecker product of two matrices takes arbitrarily shaped
  matrices. Because we are dealing with square rate matrices for QPNs,
  the Kronecker sum $\oplus$ applies to an $n^2$ matrix ${\bf A}$ and
  a $p^2$ matrix ${\bf B}$.
  \begin{center}
  ${\bf A}\otimes {\bf B}=
  \begin{bmatrix}
    a_{11}{\bf B} & \cdots & a_{1n}{\bf B} \\
    .            &     & . \\
    .            &     & . \\
    .            &     & . \\
    a_{m1}{\bf B}& \cdots & a_{mn}{\bf B}
  \end{bmatrix}
  ~~~~~~~~{\bf A}\oplus {\bf B}={\bf A}\otimes{\bf 1_p}+{\bf
    1_n}\otimes {\bf B}
  $
  \end{center}
  Note that by the assumptions above, both the Kronecker product and
  sum of the respective matrices result in square matrices for the
  required shape of the Kronecker product of QPNs.  Kronecker algebra
  has been studied intensively in the 1990s (and probably earlier) for
  the composition of stochastic Petri nets. The free juxtaposition of
  two SPNs has been shown to be the Kronecker product in terms of
  reachability graphs and their stochastic rate matrices.

  For the Kronecker product, we first note that the juxtaposition of
  two nets is a well-defined net.  Let us look at the standard
  definition of reachability for Petri nets applied to QPNs and
  reformulate this so as to show the Kronecker product formula, as
  defined in the literature both for graphs and matrices.

  $\QPN\otimes\QPN'$ in the theorem is defined as the juxtaposition
  $\QPN''=(S'',r'')$ of two isolated QPNs $\QPN=(\SysPN,r)$ and
  $\QPN'=(\SysPN',r')$, i.e., the union of the nets as graphs:
  $\PGen''=\PGen\cup\PGen'$ and $\TGen''=\TGen\cup\TGen'$ as net
  elements with the combined flow function being the union of the two
  component flow functions, each mapping the corresponding subset of
  transitions to their input and output in the respective component,
  as defined there.

  Moreover, any combination of component markings $m\in\reachset_0$
  and $m'\in\reachset'_0$ is a marking $mm'$ (as multiset product) in
  $\SysPN''$ and by def. of the QPN Kronecker product:
  $r''_{mm'}(1)=r_m(1)\times r'_{m'}(1)$.  In particular the initial
  marking is such a product marking $M''_0=M_0M'_0$.  Hence the
  cardinality of the reachability set, which is also the dimension of
  the rate square matrix, is given by
  $|\reachset''_{0}|=|\reachset_{0}|\times|\reachset'_{0}|$.  This is
  what we expect for the Kronecker product of rate matrices too.

  The combined single-transition reachability (in $S''$) includes
  $mm'\reach{t}_1m''m'$ if $m\reach{t}_1m''$ in $\QPN$ and
  $mm'\reach{t'}_1mm''$ if $m'\reach{t'}_1m''$ in $\QPN'$, with
  corresponding single transitions $t$ and $t'$.
  This defines the reachability graph $\reachgraph{S''}$, which
  is a multigraph if at least one of $\reachgraph{S}$ or $\reachgraph{S'}$
  is a multigraph. If both are simple, then so is $\reachgraph{S''}$.

  Now as for $\rategraph{Q}$, we need to consider all concurrences
  (incl. the unit concurrence $1$) implied by
  $\reachgraph{S''}$.
  \begin{enumerate}
  \item A non-unit concurrence $t\not =1$ with $m_1\reach{t}_cm_2$ in $\SysPN$ can fire
    without any firing in $\SysPN'$.  Hence $m_1m_3\reach{t}_cm_2m_3$ is
    in $\rategraph{Q''}$ for any reachable marking $m_3$ in $\SysPN'$.
    The rate for the corresponding edge in $\rategraph{Q''}$ is
    $r''_{m_1m_3}(t)=r_{m_1}(t)\times r'_{m_3}(1)$.
  \item Similarly, a non-unit concurrence $t'\not=1$ with $m_3\reach{t'}_cm_4$ in
    $\SysPN'$ can fire without any firing in $\SysPN$.  Hence
    $m_1m_3\reach{t'}_cm_1m_4$ is in $\rategraph{Q''}$ for any reachable
    marking $m_1$ in $\SysPN$.  The rate for the corresponding edge in
    $\rategraph{Q''}$ is $r''_{m_1m_3}(t')=r_{m_1}(1)\times
    r'_{m_3}(t')$.
  \item The two concurrences $t$ and $t'$ above can fire together as
    $m_1m_3\reach{tt'}_cm_2m_4$. This corresponds to an edge in
    $\rategraph{Q''}$ with rate $r''_{m_1m_3}(tt')=r_{m_1}(t)\times
    r_{m_3}(t')$.
  \item Finally any pair of unit concurrences $m\reach{1}_cm$ and
    $m'\reach{1}_cm'$ of the respective subnets can be combined, to $mm'\reach{1}_cmm'$ in $\SysPN''$
    and rated $r''_{mm'}(1)=r_m(1)\times r'_{m'}(1)$.    
  \end{enumerate}
  Again, this is what we expect of the multigraph Kronecker product
  for our QPN rate graphs, with loop edges $m\reach{1}m$ in the rate graphs, as carriers
  for unit rates $r_m(1)$.
  If $\rategraph{Q''}$ is simple--- except for the unit self-loops---,
  the terms above define exactly the Kronecker product of the
  corresponding rate matrices.  For a multigraph, the corresponding
  sums of parallel edges have to be formed, in each of the component
  rate matrices. Then, in the rate matrix of the Kronecker product,
  the corresponding product of the two sums appears correctly, because
  in Point (3) above, we have formed all corresponding pairs of edges
  for a pair of multi-edges, incl. the degenerate case of the unit concurrence $1$.

  Hence the rate matrices satisfy: $\ratematrix{Q\otimes Q'} = \ratematrix{Q''} =
  \ratematrix{Q}\otimes\ratematrix{Q'}$.

  For the Kronecker sum, we first note that the proof is
  straightforward: all the operations used in the definition of the
  Kronecker sum, i.e., units ${\bf 1}$, $+$ and $\otimes$, are
  well-defined as shown above and satisfy their own rate matrix
  equations. Consequently, the claimed Kronecker sum rate matrix
  equation $R_{Q\oplus Q'}=R_{Q}\oplus R_{Q'}$ is satisfied, too. For,
  the Kronecker sum of the two QPNs is
  $\QPN\oplus\QPN'=\QPN\otimes{\bf 1_{Q'}}+{\bf 1_{Q}}\otimes \QPN'$,
  and we can use the rate matrix equations of QPN sum and Kronecker
  product to arrive at exactly the matrix Kronecker sum shown above.

  \qed
\end{proof}
\end{document}